\documentclass[aps,prx,onecolumn,citeautoscript,nofootinbib,preprint]{revtex4}
\usepackage{amsmath}
\usepackage{amssymb}
\usepackage{float}
\usepackage[section]{placeins}
\usepackage{bbm}
\usepackage{bm}
\usepackage{comment}
\usepackage{graphicx}
\usepackage{physics}
\usepackage{color}
\usepackage[papersize={8.5in,11in}]{geometry}
\usepackage[colorlinks=true]{hyperref}
\hypersetup{
    bookmarks=true,         
    unicode=false,          
    pdftoolbar=true,        
    pdfmenubar=true,        
    pdffitwindow=false,     
    pdfstartview={FitH},    
    pdftitle={S},    
    pdfauthor={S. Sachdev},     
    pdfsubject={},   
    pdfcreator={},   
    pdfproducer={}, 
    pdfkeywords={} {} {}, 
    pdfnewwindow=true,      
    colorlinks=true,       
    linkcolor=magenta, 
    citecolor=blue,        
    filecolor=magenta,      
    urlcolor=blue           
}

\usepackage{amsfonts}
\usepackage{upgreek}
\usepackage{slashed}
\usepackage{latexsym}
\usepackage[export]{adjustbox}
\usepackage{dsfont}
\usepackage{subcaption}
\begin{document}

\preprint{\href{https://arxiv.org/abs/2305.01001}{arXiv:2305.01001}}
\title{Strange metals and black holes:\\ insights from the Sachdev-Ye-Kitaev model}
\author{Subir Sachdev}
\affiliation{Department of Physics, Harvard University, Cambridge MA-02138, USA}
\date{\today}
             \begin{abstract}
Complex many-particle quantum entanglement is a central theme in two distinct major topics in physics: the strange metal state found in numerous correlated electron compounds, and the quantum theory of black holes in Einstein gravity. The Sachdev-Ye-Kitaev model provides a solvable theory of
entangled many-particle quantum states without quasiparticle excitations. This article reviews how this toy model has led to realistic universal models of strange metals, and to new insights on the quantum states of black holes.\\~\\
Keywords: quasiparticles, strange metals, non-Fermi liquids, superconductors, black holes, entropy\\~\\
{\sffamily 
\begin{center} 
Published in the  Oxford Research Encyclopedia of Physics, December 2023,
\href{https://doi.org/10.1093/acrefore/9780190871994.013.48}{10.1093/acrefore/9780190871994.013.48} \\~\\~\\
Some of the introductory discussion in this article overlaps with the review in \href{https://arxiv.org/abs/2205.02285}{arXiv:2205.02285}.
\end{center}
}
\end{abstract}
\maketitle{}

\newpage

The last few decades have seen much progress in our understanding of the remarkable physical consequences of many-particle quantum entanglement, coming from a synthesis of ideas from quantum condensed matter theory, quantum information science, quantum field theory, and string theory. This article discusses some of the insights that have emerged from studies of the Sachdev-Ye-Kitaev (SYK) and related models, and their impact on the theory of strange metals found in numerous correlated electron compounds. Studies of this model have also had an impact on the quantum theory of black hole solutions of Einstein's theory of classical general relativity---the black hole connection is briefly noted here, and discussed further in a companion article \cite{Sachdev23}.

The author proposed a model closely related (with the same physical properties) to what is now called the SYK model  in 1992 in a paper with his first graduate student, Jinwu Ye \cite{SY}. Alexei Kitaev \cite{kitaev_talk} proposed a modification in 2015 which simplified its solution, and enabled important insights from a more refined analysis \cite{Sachdev15,kitaevsuh,Maldacena_syk,JMDS16b,Fu16,Cotler16,Bagrets17,StanfordWitten,Moitra18,Sachdev19,luca20,matt22}. The motivation in 1992 was to write down the simplest model of a metal without quasiparticles, as a starting point towards addressing the strange metal problem of the cuprates, and their marginal Fermi liquid behavior \cite{Varma89}.
Additional properties of the SYK model were described by Olivier Parcollet and Antoine Georges in Refs.~\cite{Parcollet1,GPS2} in 1999-2001, and in 2010 the author pointed out \cite{SS10,Sachdev:2010uj} that the SYK model also provided a remarkable description of the low temperature properties of certain black holes \cite{McGreevy10}. This connection has since undergone rapid development and has been made quite precise \cite{Sachdev23}. 
The SYK model shows that the quantum entanglement responsible for the absence of quasiparticles in strange metals is closely connected to that needed for a microscopic quantum theory of black holes.

\section{Foundations by Boltzmann}
\label{sec:Boltzmann}

Two foundational contributions by Boltzmann to statistical mechanics are first recalled.

In 1870, Boltzmann gave a precise definition of the  thermodynamic entropy $S$ in statistical terms:
\begin{equation}
S = k_B \ln W\,, \label{b1}
\end{equation}
where $k_B$ is Boltzmann's constant, and $W$ is number of microstates consistent with macroscopically observed properties. The value of $W$ diverges exponentially with the volume of the system, and so $S$ is extensive {\it i.e.\/} proportional to the volume. Boltzmann was thinking in terms of a dilute classical gas of molecules, as found in the atmosphere. But Boltzmann's definition works also for quantum systems, upon replacing $W$ by $D(E)$, the density of the energy eigenstates of the many-body quantum system per unit energy interval, giving
\begin{equation}
D(E) \sim \exp \left( S(E)/k_B \right)\,, \label{b2}
\end{equation}
where $S(E)$ is the thermodynamic entropy in the microcanonical ensemble with extensive energy $E$.

\begin{figure}
\begin{center}
\includegraphics[width=3in]{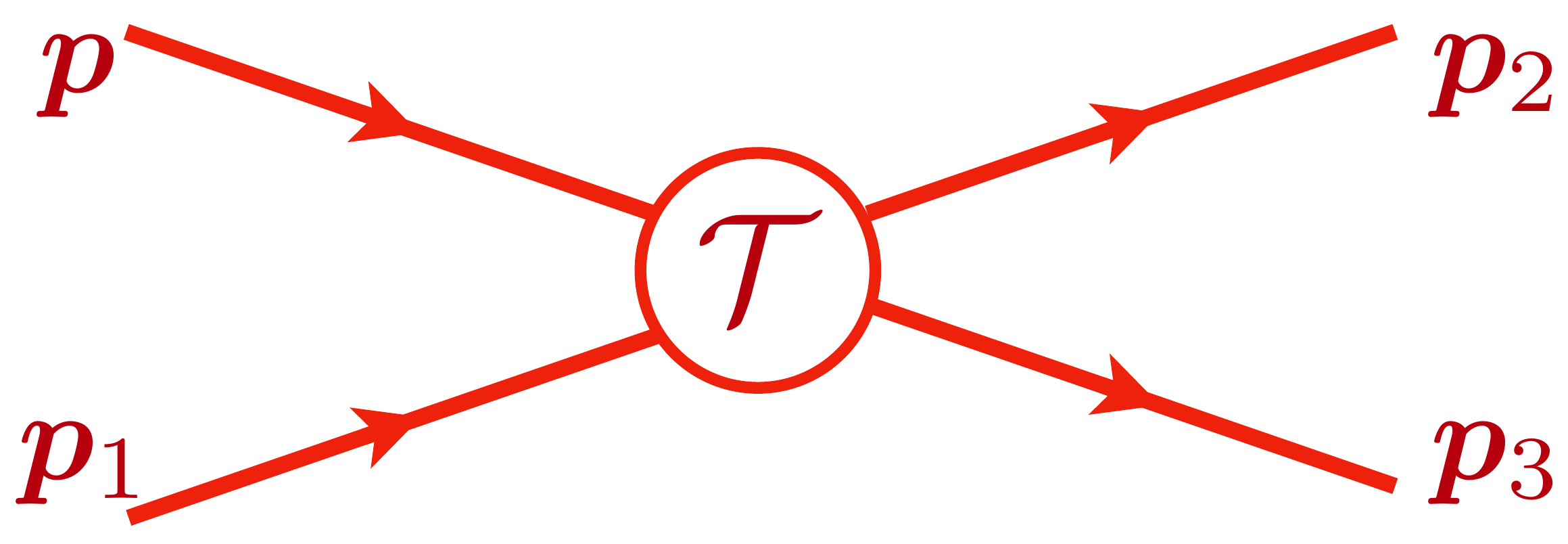}
\end{center}
\caption{Collision between two molecules. The collision term in the Boltzmann equation is proportional to the absolute square of the ${T}$-matrix.}
\label{fig1}
\end{figure}

Second, in 1872, Boltzmann's equation gave a correct description of the time evolution of the observable properties of a dilute gas in response to external forces. He applied Newton's laws of motion to individual molecules, and obtained an equation for $f_{\bm p}$, the density of particles with momentum ${\bm p}$. In a spatially uniform situation, Boltzmann's equation takes the form
\begin{equation}
\frac{\partial f_{\bm p}}{\partial t} + {\bm F} \cdot \nabla_{\bm p} f_{\bm p} = \mathcal{C}[f]\,, \label{be}
\end{equation}
where $t$ is time, and ${\bm F}$ is the external force. The left-hand-side of (\ref{be}) is just a restatement of Newton's laws for individual molecules. Boltzmann's innovation was the right-hand-side, which describes collisions between the molecules. Boltzmann introduced the concept of `molecular chaos', which asserted that in a sufficiently dilute gas successive collisions were statistically independent. With this assumption, Boltzmann showed that
\begin{equation}
\mathcal{C}[f] \propto \int_{{\bm p}_{1,2,3}} \cdots \left [f_{\bm p} f_{{\bm p}_1} - f_{{\bm p}_2} f_{{\bm p}_3} \right] \label{coll1}
\end{equation}
for a collision between molecules as shown in Fig.~\ref{fig1}. The statistical independence of collisions is reflected in the products of the densities in (\ref{coll1}), and the second term represents the time-reversed collision.

\section{Ordinary and strange metals}
\label{sec:ordinary_and_strange}

The remarkable fact is that Boltzmann's equation also applies, with relatively minor modifications, in situations very different from the dilute classical gas: it also applies to the dense quantum gas of electrons found in ordinary metals. Now collisions become rare because of Pauli's exclusion principle, and the statistical independence of collisions is assumed to continue to apply. The main modification is that the collision term in (\ref{coll1}) is replaced by
\begin{equation}
\mathcal{C}[f] \propto \int_{{\bm p}_{1,2,3}} \cdots \left [f_{\bm p} f_{{\bm p}_1}(1-f_{{\bm p}_2})(1-f_{{\bm p}_3}) - f_{{\bm p}_2} f_{{\bm p}_3} (1-f_{{\bm p}})(1-f_{{\bm p}_1})\right] \,,\label{coll2}
\end{equation}
where the additional $(1-f)$ factors ensure that the final states of collisions are not occupied. Now the $f_{\bm p}$ measure the distribution of electronic quasiparticles, and the cloud of particle-hole pairs around each electron only renormalizes the microscopic scattering cross-section.
Such a quantum Boltzmann equation is the foundation of the quasiparticle theory of the electron gas in metals, superconductors, semiconductors, and insulators, and indeed almost all of condensed matter physics before the 1980's. One of its important predictions is that as temperature $T \rightarrow 0$, the typical time between collisions, $t_c$, diverges as $t_c \sim 1/T^2$.

How short can we make $t_c$ before the quantum interference between successive collisions can no longer be ignored, and the concept of quasiparticles does not make sense? An energy-time uncertainty-principle argument indicates that {\it any\/} many-body quantum system should have a relaxation time \cite{ssbook}
\begin{equation}
t_r \geq \alpha \frac{\hbar}{k_B T} \quad, \quad T \rightarrow 0\,, \label{taueq}
\end{equation}
where $\alpha$ is a dimensionless, $T$-independent constant. For systems with quasiparticles, $t_c \sim t_r$ is expected, and the introduction of a general relaxation time $t_r$ allows a more general discussion in systems without quasiparticles. From studies of various quantum critical systems, it was argued \cite{ssbook} that the inequality in (\ref{taueq}) becomes an equality when quasiparticles are absent, as in strange metals. Recent experiments \cite{admr20} on the strange metal in cuprate superconductors have measured a particular relaxation time by connecting it to the angle dependence of the resistivity in a magnetic field, and indeed found it obeys (\ref{taueq}) as an equality, with $\alpha \approx 1.2$. This is often stated as the strange metal exhibiting `Planckian time' dynamics \cite{Hartnoll21}.

\section{The SYK model}
\label{sec:SYK}

\begin{figure}
\begin{center}
\includegraphics[width=3in]{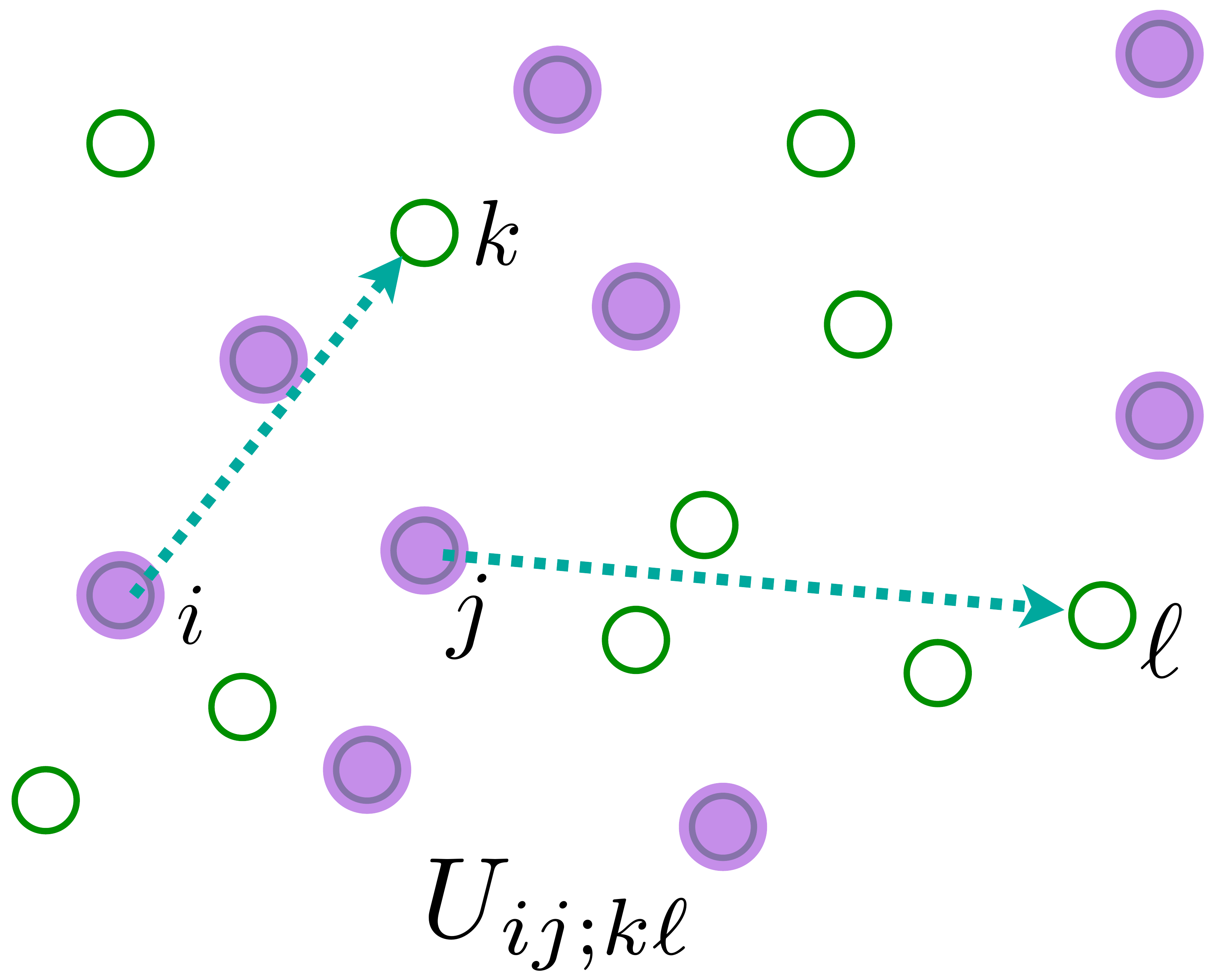}
\end{center}
\caption{The SYK model: fermions undergo the transition (`collision') shown with quantum amplitude $U_{ij;k\ell}$.}
\label{fig4}
\end{figure}

The Hamiltonian of a version of a SYK model is illustrated in Fig.~\ref{fig4}. A system with fermions $c_i$, $i=1\ldots N$ states is assumed. Depending upon physical realizations, the label $i$ could be position or an orbital, and it is best to just think of it as an abstract label of a fermionic qubit with the two states $\left|0 \right\rangle$ and $c_i^\dagger \left|0 \right\rangle$. $\mathcal{Q} N$ fermions are placed in these states, so that a density $\mathcal{Q} \approx 1/2$ is occupied, as shown in Fig.~\ref{fig4}. 
The quantum dynamics is restricted to {\it only\/} have a `collision' term between the fermions, analogous to the right-hand-side of the Boltzmann equation. However, in stark contrast to the Boltzmann equation, statistically independent collisions are not assumed, and quantum interference between successive collisions is accounted for: this is the key to building up a many-body state with non-trivial entanglement. So a collision in which fermions move from sites $i$ and $j$ to sites $k$ and $\ell$ is characterized not by a probability, but by a quantum amplitude $U_{ij;k\ell}$, which is a complex number.

The model so defined has a Hilbert space of order $2^N$ states, and a Hamiltonian determined by order $N^4$ numbers $U_{ij;k\ell}$. Determining the spectrum or dynamics of such a Hamiltonian for large $N$ seems like an impossibly formidable task. But with the assumption that the $U_{ij;k\ell}$ are statistically independent random numbers, remarkable progress is possible. Note that an ensemble of SYK models with different $U_{ij;k\ell}$ is not being considered, but a single fixed set of $U_{ij;k\ell}$. Most physical properties of this model are self-averaging at large $N$, and so as a technical tool, they can be rapidly obtained by computations on an ensemble of random $U_{ij;k\ell}$. In any case, the analytic results described below have been checked by numerical computations on a computer for a fixed set of $U_{ij;k\ell}$.
Recall that, even for the Boltzmann equation, there was an ensemble average over the initial positions and momenta of the molecules that was implicitly performed.

Specifically, the Hamiltonian in a chemical potential $\mu$ is 
\begin{align}
&\mathcal{H} = \frac{1}{(2 N)^{3/2}} \sum_{i,j,k,\ell=1}^N U_{ij;k\ell} \, c_i^\dagger c_j^\dagger c_k^{\vphantom \dagger} c_\ell^{\vphantom \dagger} 
-\mu \sum_{i} c_i^\dagger c_i^{\vphantom \dagger} \label{HH} \\
& ~~~~~~c_i c_j + c_j c_i = 0 \quad, \quad c_i^{\vphantom \dagger} c_j^\dagger + c_j^\dagger c_i^{\vphantom \dagger} = \delta_{ij}\\
&~~~~\mathcal{Q} = \frac{1}{N} \sum_i c_i^\dagger c_i^{\vphantom \dagger} \, ; \quad
[\mathcal{H}, \mathcal{Q}] = 0\, ; \quad  0 \leq \mathcal{Q} \leq 1\,,
\end{align}
and its large $N$ limit is most simply taken graphically, order-by-order in $U_{ij;k\ell}$, and averaging over $U_{ij;k\ell}$ as independent random variables with $\overline{U_{ij;k\ell}} = 0$ and $\overline{|U_{ij;k\ell}|^2} = U^2$.
\begin{figure}
\begin{center}
\includegraphics[width=3in]{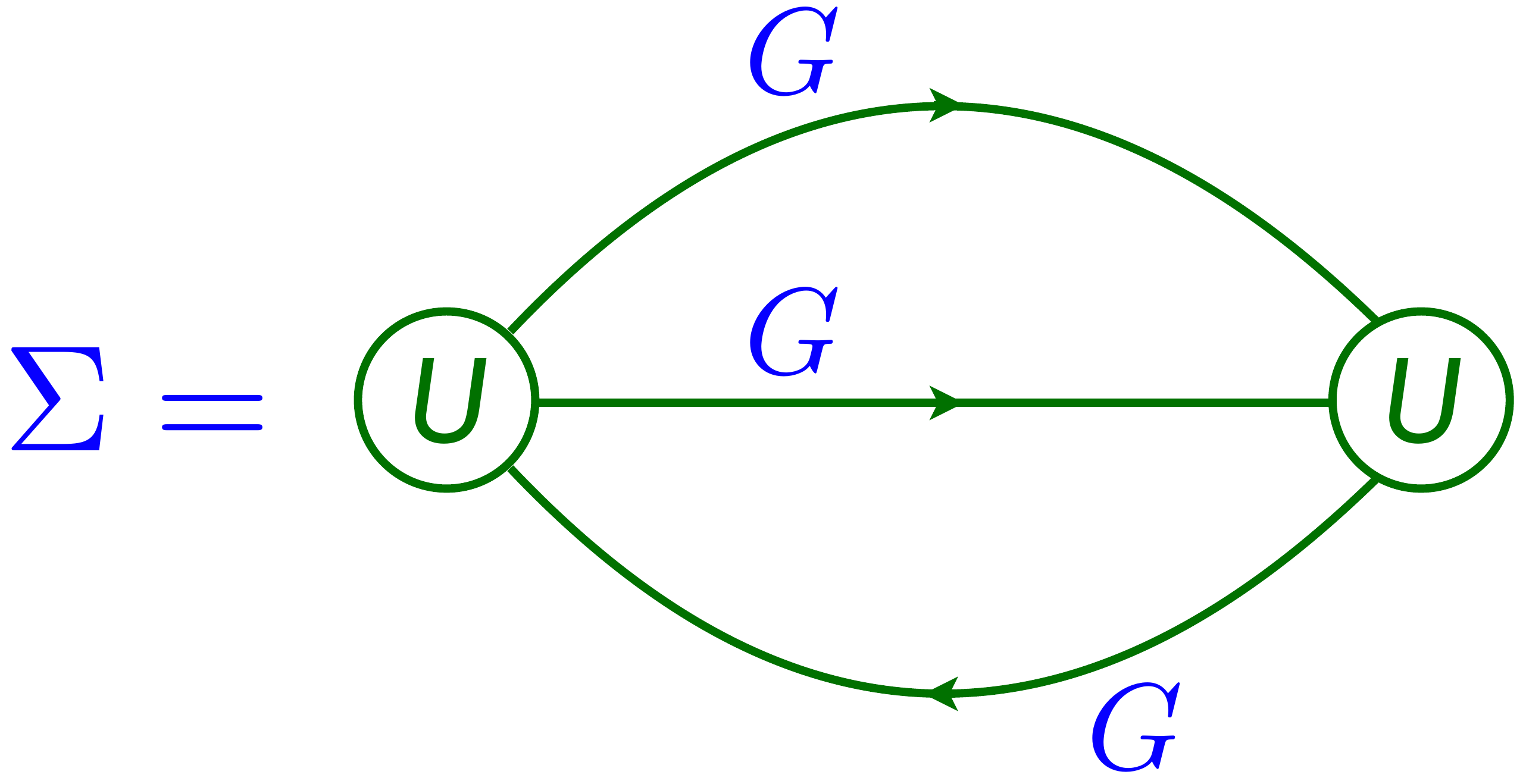}
\end{center}
\caption{Self-energy for the fermions of $\mathcal{H}$ in (\ref{HH}) in the limit of large $N$. The intermediate Green's functions are fully renormalized.}
\label{fig:sygraph}
\end{figure}
This expansion can be used to compute graphically the Green's function in imaginary time $\tau$
\begin{align}
G(\tau) = - \frac{1}{N} \sum_i \overline{\left\langle \mathcal{T} \left( c_i (\tau) c_i^\dagger (0) \right) \right\rangle}\,, \label{G1}
\end{align}
where $\mathcal{T}$ is the time-ordering symbol, the angular brackets are a quantum average for any given $U_{ij;k\ell}$, and the over-line denotes an average over the ensemble of $U_{ij;k \ell}$. (It turns out that  the last average is not needed for large $N$, because the quantum observable is self-averaging.)
In the large $N$ limit, only the graph for the Dyson self energy, $\Sigma$, in Fig.~\ref{fig:sygraph} survives, and the on-site fermion Green's function is given by the solution of the following equations
\begin{align}
G(i\omega_n) &= \frac{1}{i \omega_n + \mu - \Sigma (i\omega_n)} \nonumber \\ 
\Sigma (\tau) & = -  U^2 G^2 (\tau) G(-\tau) \nonumber \\
G(\tau = 0^-) & = \mathcal{Q}\,, \label{sy1}
\end{align}
where  $\omega_n$ is a fermionic Matsubara frequency. The first equation in (\ref{sy1}) is the usual Dyson relation between the Green's function and self energy in quantum field theory, the second equation in (\ref{sy1}) is the Feynman graph in Fig.~\ref{fig:sygraph}, and the last determines the chemical potential $\mu$ from the charge density  $\mathcal{Q}$.  These equations can also be obtained as saddle-point equations of the following exact representation of the disordered-averaged partition function, expressed as a `$G-\Sigma$' theory \cite{GPS2,Sachdev15,kitaevsuh,Maldacena_syk}:
\begin{align}
\mathcal{Z} &= \int \mathcal{D} G(\tau_1, \tau_2) \mathcal{D} \Sigma (\tau_1, \tau_2) \exp (-N I) \nonumber \\
I &= \ln \det \left[ \delta(\tau_1 - \tau_2) (\partial_{\tau_1} + \mu) - \Sigma (\tau_1, \tau_2) \right] \nonumber \\
&~~~
+ \int d \tau_1 d \tau_2  \left[ \Sigma(\tau_1, \tau_2) G(\tau_2, \tau_1) + (U^2/2) G^2(\tau_2, \tau_1) G^2(\tau_1, \tau_2) \right] \label{GSigma1}
\end{align}
This is a path-integral over bi-local in time functions $G(\tau_1, \tau_2)$ and $\Sigma(\tau_1, \tau_2)$, whose saddle point values are the Green's function $G(\tau_1 - \tau_2)$,  and the self energy $\Sigma (\tau_1 - \tau_2)$. This bi-local $G$ can be viewed as a composite quantum operator corresponding to an on-site fermion bilinear
\begin{align}
G(\tau_1, \tau_2) = - \frac{1}{N} \sum_i  \mathcal{T} \left( c_i (\tau_1) c_i^\dagger (\tau_2) \right) 
\end{align}
that is averaged in (\ref{G1}).

For general $\omega$ and $T$, the equations in (\ref{sy1}) have to be solved numerically. But an exact analytic solution is possible in the limit $\omega, T \ll U$. 
At $T=0$, the asymptotic forms can be obtained straightforwardly \cite{SY}
\begin{align}
G(i \omega) \sim -i \mbox{sgn} (\omega) |\omega|^{-1/2} \quad, \quad \Sigma(i \omega) - \Sigma (0) \sim -i \mbox{sgn} (\omega) |\omega|^{1/2}\,,
\label{sy10}
\end{align}
and a more complete analysis of (\ref{sy1}) gives the exact form at non-zero $T$ ($\hbar = k_B = 1$) \cite{Parcollet1}
\begin{align}
G (\omega)  = \frac{-i C e^{-i \theta}}{(2 \pi T)^{1/2}}
\frac{\Gamma \left( \displaystyle \frac{1}{4} - \frac{i  \omega}{2 \pi T} + i \mathcal{E} \right)}
{\Gamma \left(  \displaystyle \frac{3}{4} - \frac{i \omega }{2 \pi T} + i \mathcal{E} \right)} \quad\quad  |\omega|, T \ll U \,. \label{sy2}
\end{align}
Here, $\mathcal{E}$ is a dimensionless number which characterizes the particle-hole asymmetry of the spectral function; both $\mathcal{E}$ and the pre-factor $C$ are determined by an angle $-\pi/4 < \theta < \pi/4$
\begin{align}
e^{2 \pi \mathcal{E}} = \frac{\sin(\pi/4 + \theta)}{\sin(\pi/4 - \theta)} \quad, \quad  C = \left( \frac{\pi}{U^2 \cos (2 \theta) }\right)^{1/4}\,,
\end{align}
and the value of $\theta$ is determined by a Luttinger relation to the density $\mathcal{Q}$ \cite{GPS2}
\begin{align}
\mathcal{Q} = \frac{1}{2} - \frac{\theta}{\pi} - \frac{\sin(2 \theta)}{4}\,.
\end{align}

A notable property of (\ref{sy2}) at $\mathcal{E}=0$ is that it equals the temporal Fourier transform of the spatially local correlator of a fermionic field of dimension 1/4 in a conformal field theory in 1+1 spacetime dimensions. A theory in 0+1 dimensions is considered here, where conformal transformations map the temporal circle onto itself, as reviewed in Appendices A and B of Ref.~\cite{SYKRMP}; such transformations allow a non-zero $\mathcal{E}$. An important consequence of this conformal invariance is that (\ref{sy2}) is a scaling function of $\hbar \omega/(k_B T)$ (after restoring fundamental constants); in other words, the characteristic frequency scale of (\ref{sy2}) is determined solely by $k_B T/\hbar$, is independent of the value of $U/\hbar$, and saturates the bound in (\ref{taueq}). A careful study of the consequences of this conformal invariance have established the following properties of the SYK model (more complete references to the literature are given in other reviews \cite{SYKRMP,QPMbook}):
\begin{itemize} 
\item There are no quasiparticle excitations, and the SYK model exhibits quantum dynamics with a Planckian relaxation time obeying (\ref{tra}) at $T \ll U$. In particular, the relaxation time is {\it independent\/} of $U$, a feature not present in any ordinary metal with quasiparticles. While the Planckian relaxation in (\ref{sy2}) implies the absence of quasiparticles with the same quantum numbers as the $c$ fermion, it does not rule out the possibility that $c$ has fractionalized into some emergent quasiparticles; this possibility is ruled out by the exponentially large number of low energy states, as discussed below.
\item At large $N$, the many-body density of states at fixed $\mathcal{Q}$ is \cite{Cotler16,Bagrets17,Maldacena_syk,kitaevsuh,StanfordWitten,GKST} (see Fig.~\ref{fig5}a)
\begin{equation}
D(E) \sim \frac{1}{N} \exp (N s_0) \sinh \left( \sqrt{2 N \gamma E} \right)\,, \label{de}
\end{equation}
where the ground state energy has been set to zero.
Here $s_0$ is a universal number dependent only on $\mathcal{Q}$ ($s_0 = 0.4648476991708051 \ldots$ for $\mathcal{Q}=1/2$), $\gamma \sim 1/U$ is the only parameter dependent upon the strength of the interactions, and the $N$ dependence of the pre-factor is discussed in Ref.~\cite{GKST}.
\begin{figure}
\begin{center}
\includegraphics[width=6in]{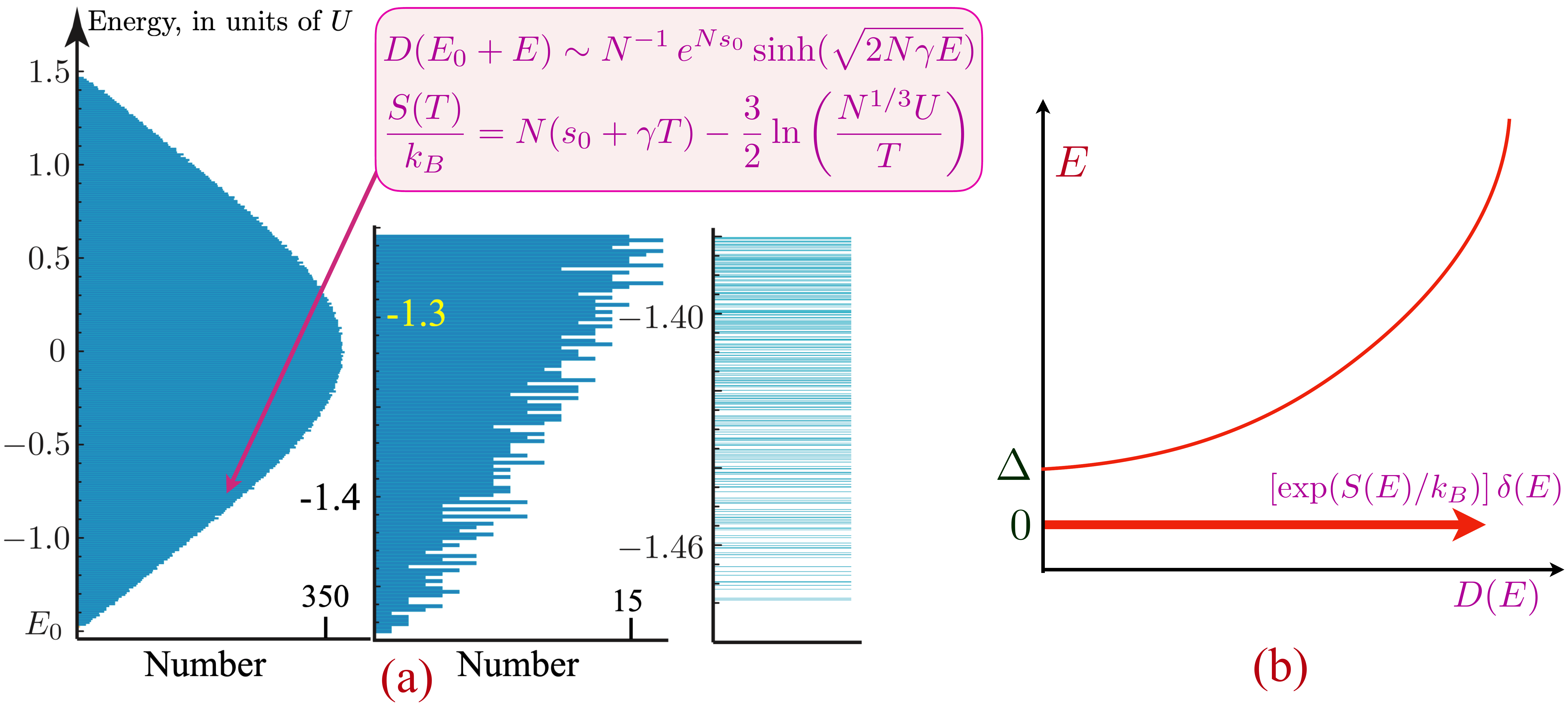}
\end{center}
\caption{(a) Plot of the 65536 many-body eigenvalues of a $N = 32$ Majorana SYK Hamiltonian; however, the analytical results quoted here are for the SYK model with complex fermions which has a similar spectrum. The coarse-grained low-energy and 
low-temperature behavior is described by (\ref{de}) and (\ref{SSYK}). 
(b) Schematic of the lower energy density of states of a supersymmetric generalization of the SYK model \cite{Fu16,StanfordWitten}. There is a delta function at $E=0$, and
the energy gap $\Delta$ is proportional to the inverse of $S(E=0)$.}
\label{fig5}
\end{figure}
Given $D(E)$, the partition function can be computed from (\ref{ZD}) at a temperature $T$, and hence the low-$T$ dependence of the entropy at fixed $Q$ is given by
\begin{equation}
\frac{S(T)}{k_B} = N(s_0 + \gamma \, k_B T) - \frac{3}{2}\ln \left(\frac{U}{k_B T} \right) - \frac{\ln N}{2} + \ldots \,. \label{SSYK}
\end{equation}
The thermodynamic limit $\lim_{N \rightarrow \infty} S(T)/N$ yields the microcanonical entropy 
\begin{align}
S(E)/k_B = Ns_0 + \sqrt{2N \gamma E}\,, 
\end{align}
and this connects to the extensive $E$ limit of (\ref{de}) after using Boltzmann's formula in (\ref{b2}).
The limit $\lim_{T \rightarrow 0} \lim_{N \rightarrow \infty} S(T)/(k_B N) = s_0$ is non-zero, implying an energy-level spacing exponentially small in $N$ near the ground state: the density of states (\ref{de}) implies that any small energy interval near the ground state contains an exponentially large number of energy eigenstates (see Fig.~\ref{fig5}a).
This is very different from systems with quasiparticle excitations, whose energy level spacing vanishes with a positive power of $1/N$ near the ground state, as quasiparticles have order $N$ quantum numbers. The exponentially small level spacing therefore rules out the existence of quasiparticles in the SYK model. 
\item
However, it important to note that there is no exponentially large degeneracy of the ground state itself in the SYK model, unlike that in a supersymmetric generalization of the SYK model (see Fig.~\ref{fig5}b) and the ground states in Pauling's model of ice \cite{Pauling}. Obtaining the ground-state degeneracy requires the opposite order of limits between $T$ and $N$, and numerical studies show that the entropy density does vanish in such a limit for the SYK model.  The many-particle wavefunctions of the 
low-energy eigenstates in Fock space change chaotically from one state to the next, providing a realization of maximal many-body quantum chaos \cite{Maldacena16} in a precise sense.
This structure of eigenstates is very different from systems with quasiparticles, for which the lowest energy eigenstates differ only by adding and removing a few quasiparticles. 
\item
The $E$ dependence of the density of states in (\ref{de}) is associated with a time reparameterization mode, and (\ref{de}) shows that its effects are important when $E \sim 1/N$. The low energy quantum fluctuations of (\ref{GSigma1}) can be expressed in terms of a path integral which reparameterizes imaginary time $\tau \rightarrow f(\tau)$, in a manner analogous to the quantum theory of gravity being expressed in terms of the fluctuations of the spacetime metric. There are also quantum fluctuations of a phase mode $\phi (\tau)$, whose time derivative is the charge density, and the path integral in (\ref{GSigma1}) reduces to the partition function
\begin{equation}
\mathcal{Z}_{SYK-TR} = e^{N s_0} \int \mathcal{D} f \mathcal{D} \phi \exp \left( - \frac{1}{\hbar} \int_0^{\hbar/(k_B T)}\!\!\!\!\!\!  d \tau \, \mathcal{L}_{SYK-TR} [ f,\phi] \right) \label{feynsyk}
\end{equation}
The Lagrangian $\mathcal{L}_{SYK-TR}$ is known, and involves a Schwarzian of $f(\tau)$. Remarkably, despite its non-quadratic Lagrangian, the path integral in (\ref{feynsyk}) can be performed exactly \cite{StanfordWitten}, and leads to (\ref{de}).
\end{itemize}

\subsection{The Yukawa-SYK model}
\label{YSYK}

The SYK model defined above is a 0+1 dimensional theory with no spatial structure, and so cannot be directly applied to transport of strange metals in non-zero spatial dimensions. A great deal of work has been undertaken on generalizing the SYK model to non-zero spatial dimensions \cite{SYKRMP}, but this effort has ultimately not been successful: although `bad metal' states (see Section~\ref{sec:strange}) have been obtained, low $T$ strange metals have not. But another effort based upon a variation of the SYK model, the 0+1 dimensional `Yukawa-SYK' model \cite{Fu16,Murugan:2017eto,Patel:2018zpy,Marcus:2018tsr,Wang:2019bpd,Ilya1,Wang:2020dtj,Altman1,WangMeng21,Schmalian1,Schmalian2,Schmalian3}, has been a much better starting point for a non-zero spatial dimensional theory, as shown in Section~\ref{sec:strange}. The present subsection describes the basic properties of the simplest realization \cite{Ilya1,Schmalian1,Schmalian2,Schmalian3} of the Yukawa-SYK model.

\begin{figure}
\begin{center}
\includegraphics[width=2.25in]{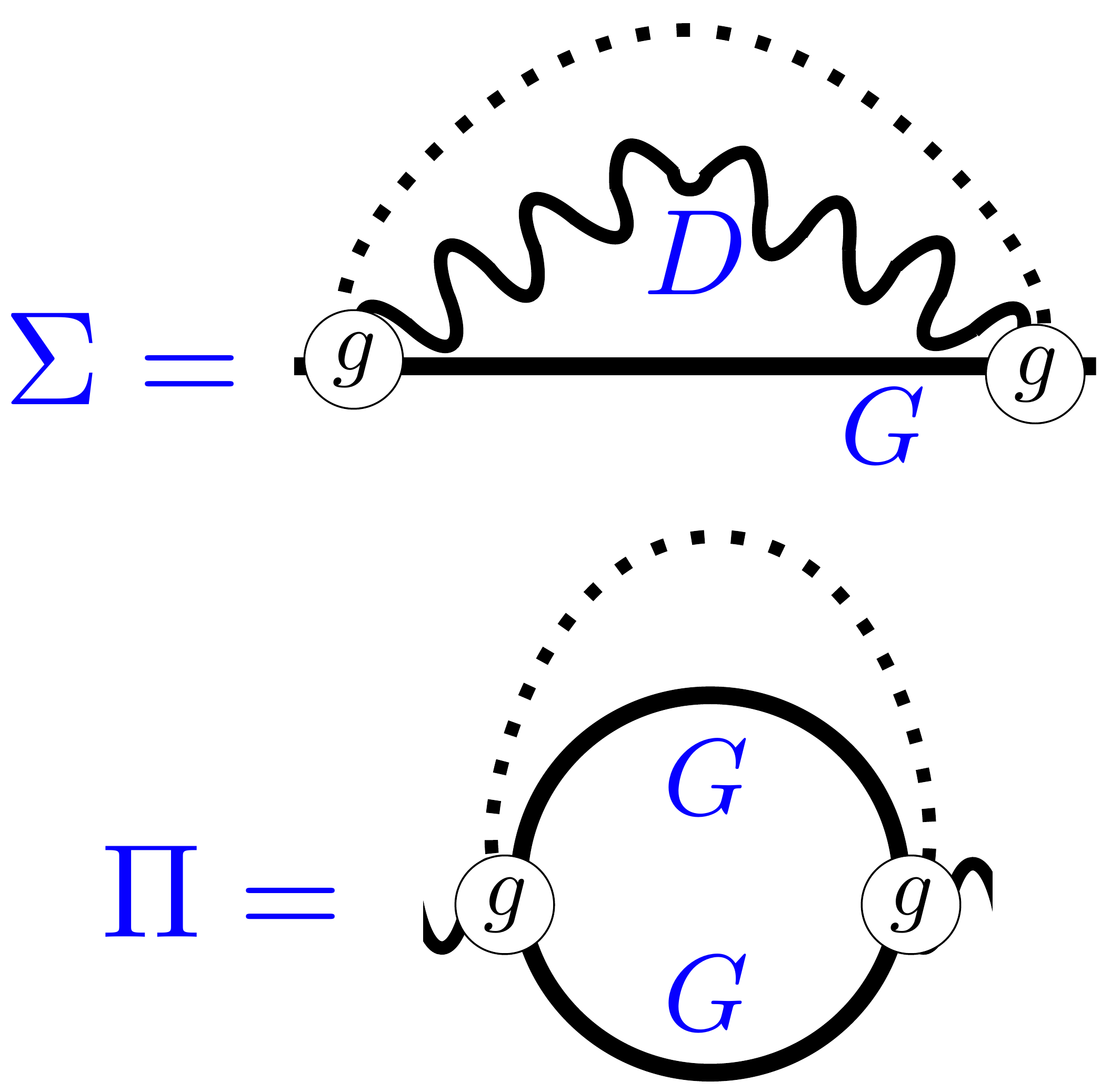}
\end{center}
\caption{Self-energies of the fermions and bosons in the Hamiltonian $\mathcal{H}_Y$ in (\ref{HY}). The intermediate Green's functions are fully renormalized.}
\label{fig:yukawa}
\end{figure}

In the spirit of (\ref{HH}), a model of fermions $c_i$ ($i=1 \ldots N$) and bosons $\phi_\ell$ ($\ell = 1 \ldots N$) with a Yukawa coupling $g_{ij\ell}$ between them is now considered
\begin{align}
\mathcal{H}_Y = -\mu \sum_{i} c_i^\dagger c_i^{\vphantom\dagger} + \sum_{\ell} \frac{1}{2} \left( \pi_\ell^2 + \omega_0^2 \phi_\ell^2 \right) + \frac{1}{N}\sum_{ij\ell} g_{ij\ell}^{\vphantom\dagger} c_i^\dagger c_j^{\vphantom\dagger} \phi_\ell^{\vphantom\dagger} \,,\label{HY}
\end{align}
with $g_{ij\ell}$ independent random numbers with zero mean and r.m.s. value $g$. The bosons are oscillators with the same frequency $\omega_0$, while the fermions have no one-particle hopping. The large $N$ limit of (\ref{HY}) can be taken just as for the SYK model in (\ref{HH}). The self-energy graph in Fig.~\ref{fig:sygraph} is replaced by those in Fig.~\ref{fig:yukawa}: the phonon Green's function is $D$, while the phonon self-energy is $\Pi$.

Continuing the parallel with the SYK model, 
the disorder-averaged partition function of the Yukawa-SYK model is a bi-local $G$-$\Sigma$-$D$-$\Pi$ theory, analogous to (\ref{GSigma1}):
\begin{equation}\label{Sall}
  \begin{split}
  \mathcal{Z} & = \int \mathcal{D} G \, \mathcal{D} \Sigma \, \mathcal{D} D \, \mathcal{D} \Pi \exp( - N S_{\rm all}) \\
    S_{\rm all} & = -\ln\det(\partial_\tau-\mu+\Sigma)+\frac{1}{2}\ln\det(-\partial_\tau^2+\omega_0^2-\Pi) \\
       & +\int d\tau  \int d \tau' \left[- \Sigma(\tau';\tau)G(\tau,\tau')+\frac{1}{2}\Pi(\tau',\tau)D(\tau,\tau') + \frac{g^2}{2}G(\tau,\tau')G(\tau',\tau)D(\tau,\tau')   \right]\,.
  \end{split} 
\end{equation}
The large $N$ saddle-point equations replacing (\ref{sy1}) are:
\begin{align}
G(i \omega_n) = \frac{1}{i \omega_n + \mu - \Sigma (i \omega_n)} \quad &, \quad D(i \omega_n) = \frac{1}{\omega_n^2 + \omega_0^2 - \Pi (i \omega_n)} \nonumber \\
\Sigma (\tau) = g^2 G(\tau) D(\tau) \quad &, \quad \Pi (\tau) = - g^2 G(\tau) G(-\tau) \label{ysyk1}
\end{align}

The solution of (\ref{Sall}) and (\ref{ysyk1}) leads to a critical state with properties very similar to that of the SYK model \cite{Ilya1,Schmalian1,Schmalian2,Schmalian3}. Only the low-frequency behavior of the Green's functions at $T=0$, is quoted analogous to (\ref{sy10}):
\begin{align}
G(i \omega) \sim -i \mbox{sgn} (\omega) |\omega|^{-(1-2 \Delta)} \quad, \quad D(i \omega) \sim |\omega|^{1-4 \Delta} \quad , \quad \frac{1}{4} < \Delta < \frac{1}{2}\,. \label{ysyk10}
\end{align}
Inserting the ansatz (\ref{ysyk10}) into (\ref{ysyk1}) fixes the value of the critical exponent $\Delta$. 
\begin{align}
\frac{4 \Delta - 1}{2(2 \Delta - 1) [ \sec(2 \pi \Delta) - 1 ]} = 1 \quad , \quad \Delta = 0.42037 \ldots \label{Deltaval}
\end{align}
Although the fermion Green's function has an exponent which differs from that of the SYK model, the thermodynamic properties have the same structure as that of the SYK model, including the presence of the Schwarzian mode and the form of the many-body density of states.

\section{From the SYK model to strange metals}
\label{sec:strange}

Some of the key properties of strange metals, as observed in recent experiments, are first summarized \cite{Hartnoll21}:
\begin{enumerate}
\item The resistivity, $\rho (T)$, of strange metals has a linear-$T$ dependence at low temperatures:
\begin{align}
\rho(T) = \rho_0 + A T + \ldots \quad, \quad T \rightarrow 0. \label{p1}
\end{align}
Importantly, this resistivity is below the Mott-Ioffe-Regel bound \cite{Hartnoll21}, so $\rho (T) < h/e^2$ in $d=2$ spatial dimensions.
Metals with $\rho (T) > h/e^2$ are {\it bad\/} metals, and are not discussed here. 
Bad metals can be described by lattice models of coupled SYK `islands' \cite{SongBalents,PatelArovas,ChowdhuryBerg}, as reviewed elsewhere \cite{SYKRMP}.
\item Ordinary metals have low $T$ specific heat which vanishes linearly with $T$, but in a strange metal the 
specific heat is enhanced to $\sim T \ln (1/T)$ as $T \rightarrow 0$.
\item Careful analyses of optical data in the cuprates over wide ranges of frequencies and temperatures \cite{NormanChubukov,Michon22} has shown that the optical conductivity can be accurately described by the following form
\begin{align}
\sigma (\omega ) = \frac{K}{\displaystyle \frac{1}{\tau_{\rm trans} (\omega)} - i \omega \frac{m_{\rm trans}^\ast (\omega)}{m}} \quad; \quad \frac{1}{\tau_{\rm trans} (\omega)} \sim |\omega| \Phi_\sigma \left( \frac{\hbar \omega}{k_B T} \right)\,, \label{p2}
\end{align}
where $K$ is a constant, and the transport scattering rate $1/\tau_{\rm trans}$ scales linearly with the larger of $|\omega|$ and $k_B T/\hbar$. The frequency dependence of the effective transport mass $m_{\rm trans}^\ast$ is then determined by a Kramers-Kronig connection to that of $1/\tau_{\rm trans}$, which leads to a logarithmic frequency dependence in $m_{\rm trans}^\ast (\omega) $. 
\item 
Photoemission experiments on the cuprates have measured the electron self-energy near the nodal point in the Brillouin zone. This was found to obey the scaling form \cite{Reber2019}
\begin{align}
\frac{1}{\tau_{\rm in} (\omega)} = 2 \, \mbox{Im} \Sigma (\omega) \sim |\omega|^{2\alpha} \Phi_{\Sigma} \left( \frac{\hbar \omega}{k_B T} \right) \label{p3}
\end{align}
with an exponent $\alpha \approx 1/2$ near optimal doping. The value $\alpha=1/2$ corresponds to a `marginal Fermi liquid' \cite{Varma89}, at least as far as the self energy is concerned. But an important point is that there is no direct theoretical connection between the single-particle scattering rate $1/\tau_{\rm in} (\omega)$ in (\ref{p3}), and the value of transport scattering rate  $1/\tau_{\rm trans}(\omega)$ in (\ref{p2}), although they are observed to have the same exponent. As seen below, the transport and single-particle scattering rates can be very different in some common models.
\item In experimental observations \cite{Bruin13,Legros19,Gael21}, the value of the overall constant $K$ in (\ref{p2}) is often fixed by writing the d.c. conductivity in the Drude form
\begin{align}
\sigma = \frac{ n e^2 \tau_{\rm trans}}{m^\ast}\,,
\label{eq:Drude}
\end{align}
where $n$ is the known conduction electron density, and $m^\ast$ is an electronic effective mass. In some experiments, the transport mass $m_{\rm trans}^\ast$ of (\ref{p2}) is used in (\ref{eq:Drude}), while other experiments use the $m^\ast$ determined from thermodynamic measurements. In the form (\ref{eq:Drude}), the absolute value of $\tau_{\rm trans}$ can be deduced from experimental observations. In the strange metal, such a value is found to obey `Planckian' behavior with \cite{Bruin13,Legros19,Gael21}
\begin{align}
\frac{1}{\tau_{\rm trans}} = \alpha \, \frac{k_B T}{\hbar}\,, \label{Planckian}
\end{align}
\end{enumerate}
with $\alpha$ a numerical constant of order unity. Measurements of $1/\tau_{\rm trans}$ in La$_{1.6-x}$Nd$_{0.4}$Sr$_x$CuO$_4$ in angle-dependent magnetotransport show $\alpha = 1.2 \pm 0.4$ \cite{Gael21} upon using the thermodynamic $m^\ast$.

\subsection{Universal model}
\label{sec:univ}

This subsection will present a simple and universal generalization of the Yukawa-SYK model of Section~\ref{YSYK} to spatial dimension $d=2$ which reproduces all five of the above observed properties \cite{Patel2}.

First, the fermions are given a two-dimensional momentum ${\bm k}$, and an associated dispersion $\varepsilon ({\bm k})$ so that the fermionic Lagrangian is
\begin{align}
\mathcal{L}_c = c_{{\bm k}}^\dagger (\tau) \left( \frac{\partial}{\partial \tau} + \varepsilon ({\bm k})\right) c_{\bm k} (\tau). \label{f1}
\end{align}
We need metallic behavior, so there is a Fermi surface where $\varepsilon ({\bm k}) = 0$. The fermions $c_{\bm k}$ may also have additional spin or flavor labels, but these are omitted for simplicity.

\begin{figure}
\begin{center}
\includegraphics[width=5in]{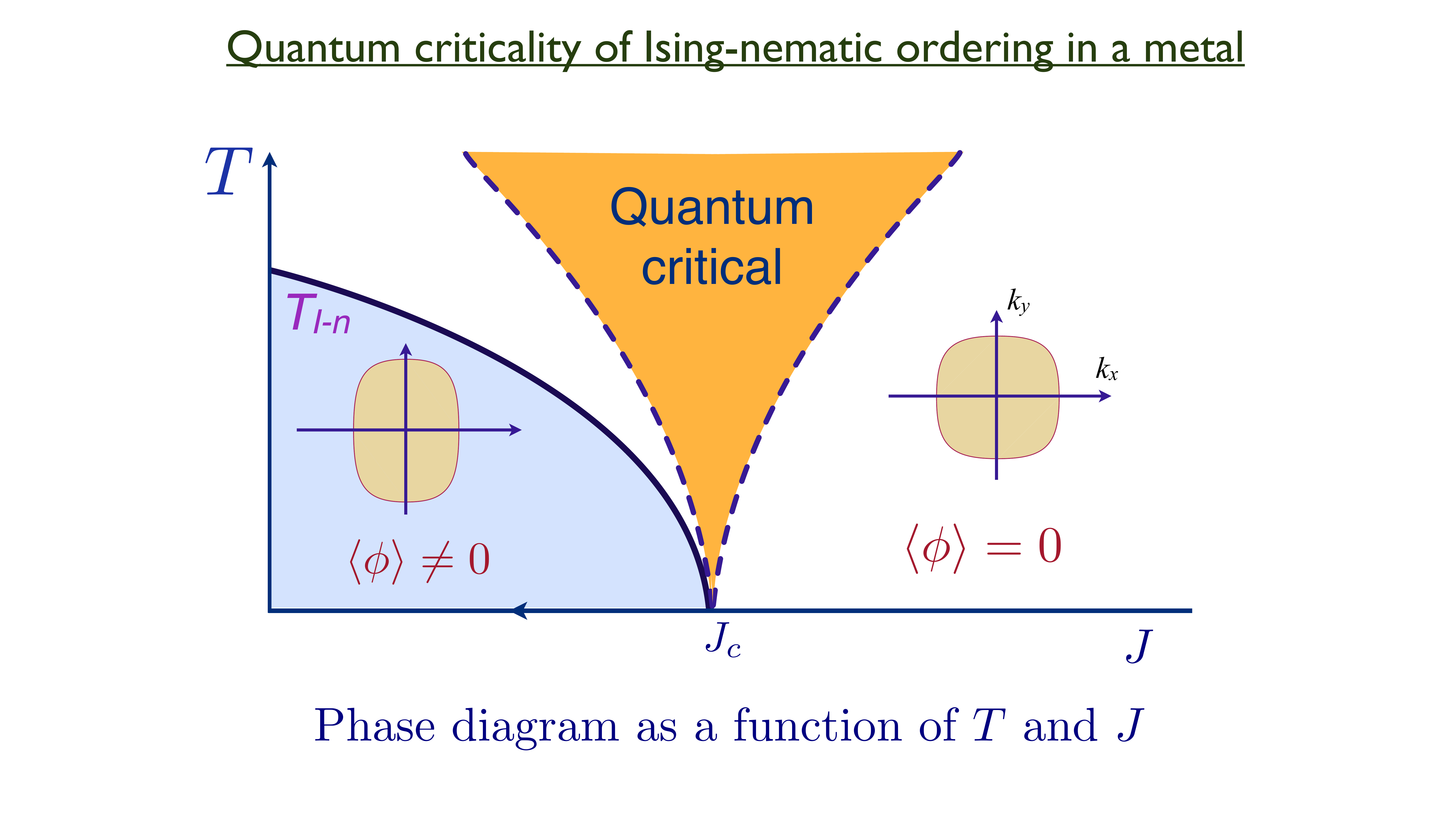}
\end{center}
\caption{Schematic phase diagram of a metal with an Ising-nematic transition. The symmetry-broken phase with $\langle \phi \rangle \neq 0$ breaks the square lattice symmetry of the Fermi surface to that of a rectangular lattice. The quantum-critical region can be present even when the $T=0$ quantum critical point is `hidden' by the onset of other phases ({\it e.g.\/} superconductivity) at low temperatures.}
\label{fig:nematic}
\end{figure}

We will obtain the scalar field $\phi$ by decoupling the electron-electron interactions in some suitable channel. Thus, schematically,  a Hubbard-Stratonovich transformation \cite{QPMbook} is performed on the quartic fermion term (all form-factors are omitted from this discussion)
\begin{align}
-J   \, c^{\dagger} ({\bm r},\tau)  c^\dagger ({\bm r},\tau) c ({\bm r},\tau)  c ({\bm r},\tau)\,, \label{f2}
\end{align}
to obtain a scalar field with a Yukawa coupling to the fermions
\begin{align}
 \frac{[\phi({\bm r}, \tau)]^2}{J} +   \, c^{\dagger}({\bm r},\tau)  c({\bm r},\tau) \, \phi({\bm r},\tau) \,. \label{f3}
\end{align}
As a specific example, consider the Ising-nematic transition of a Fermi liquid, as illustrated in Fig.~\ref{fig:nematic}.
Here $\phi$ is a measure of the distortion of the Fermi surface from perfect square lattice symmetry. Quantum fluctuations of $\phi$ 
near the quantum phase transition to the phase in which the square lattice symmetry is spontaneously broken are particularly interesting. 
Other possible realizations of $\phi$ are discussed in Section~\ref{sec:kondo}, including cases where $\phi$ does not represent a symmetry-breaking order parameter.

The discussion above has considered spatially uniform metals in which there is perfect translational symmetry, and total momentum is conserved (apart from rare umklapp events).
However, as shown below, the transport properties of such a clean metal are very far from those of the strange metals described at the beginning of Section~\ref{sec:strange}, even in the quantum critical regime of Fig.~\ref{fig:nematic}. For the universal model described here, spatial disorder turns out to be essential. The most common form of spatial disorder in studies of metallic transport is a random potential from impurities
\begin{align}
\mathcal{L}_{v} =  v({\bm r}) c^\dagger ({\bm r},\tau) c  ({\bm r},\tau)\,, \label{f4}
\end{align}
where $v({\bm r})$ is a fixed random function of position. This spatial disorder is averaged over after assuming it is uncorrelated at different points in space
\begin{align}
\overline{v ({\bm r})} = 0 \quad, \quad \overline{v({\bm r}) v({\bm r'})} = v^2 \delta({\bm r}-{\bm r'})\,. \label{f5}
\end{align}
One of the main new points made in recent work \cite{Guo2022,Patel2} is that random potential disorder is also not sufficient to produce a strange metal, and the effects of spatial randomness in the interactions \cite{Altman1} must also be considered. Specifically, the coupling $J$ in (\ref{f3}) should be allowed to also have a spatially random component
$J \rightarrow J + J' ({\bm r})$, where $J' ({\bm r})$ is a fixed random variable obeying averages similar to (\ref{f5}). 
After re-scaling $\phi$, the randomness can be transferred to the Yukawa coupling, leading finally to the following Lagrangian for the scalar field $\phi$
\begin{align}
\mathcal{L}_\phi = \frac{1}{2} \left[ \left( \partial_\tau \phi ({\bm r}, \tau) \right)^2 + \left( {\bm \nabla}_{\bm r} \phi ({\bm r}, \tau) \right)^2  + m_b^2 \left(\phi ({\bm r}, \tau) \right)^2 \right]
+ [g +  g'({\bm r})] \, c^{\dagger}({\bm r},\tau)  c({\bm r},\tau) \, \phi({\bm r},\tau)\,, \label{f6}
\end{align}
where $m_b$ is a boson mass which tunes the system to criticality, the Yukawa coupling $g$ is spatially uniform, while the coupling $g'({\bm r})$ is spatially random and obeys the disorder averages
\begin{align}
\overline{g'({\bm r})} = 0 \quad, \quad \overline{g'({\bm r}) g'({\bm r'})} = g'^2 \delta({\bm r}-{\bm r'}). \label{f7}
\end{align}
The Lagrangian $\mathcal{L}_c + \mathcal{L}_v + \mathcal{L}_\phi$ in (\ref{f1},\ref{f4},\ref{f6})  is the universal theory of a two-dimensional strange metal, generalizing the Yukawa-SYK Hamiltonian in (\ref{HY}).

\begin{figure}
\begin{center}
\includegraphics[width=3in]{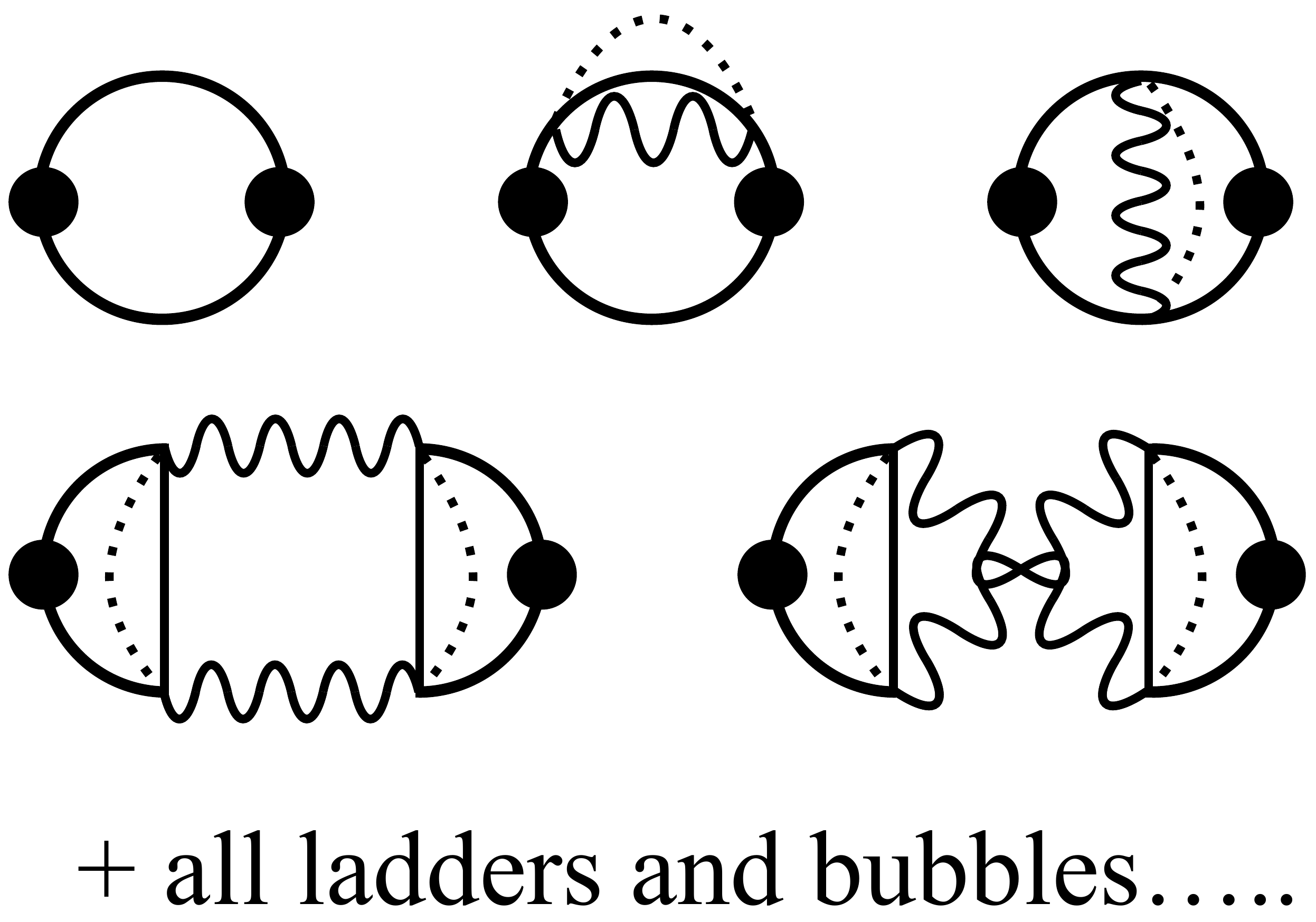}
\end{center}
\caption{Diagrams for the conductivity for the theory $\mathcal{L}_c + \mathcal{L}_v + \mathcal{L}_\phi$.}
\label{fig:ladders}
\end{figure}

The properties of this strange metal theory will be determined by directly extending the methods used to solve the Yukawa-SYK model. This extension can be viewed as simply solving the equations in Fig.~\ref{fig:yukawa}, while using the propagators in $\mathcal{L}_c + \mathcal{L}_v + \mathcal{L}_\phi$. Alternatively, a fictitious flavor index on all fields ranging over $N$ values can be introduced and the large $N$ limit taken, assuming couplings are random in this flavor space. This method yields a 
 $G$-$\Sigma$-$D$-$\Pi$ theory which is a direct generalization of (\ref{Sall}) to Green's functions that are bilocal in {\it both\/} space and time
\begin{equation}\label{Sall2}
  \begin{split}
  \mathcal{Z} & = \int \mathcal{D} G \, \mathcal{D} \Sigma \, \mathcal{D} D \, \mathcal{D} \Pi \exp( - N S_{\rm all}) \\
    S_{\rm all} & = -\ln\det(\partial_\tau+\varepsilon({{\bm k}})-\mu+\Sigma)+\frac{1}{2}\ln\det(-\partial_\tau^2+{{\bm q}}^2+m_b^2-\Pi) \\
       & +\int d\tau d^2 r \int d \tau' d^2 r'\left[- \Sigma(\tau',{\bm r}';\tau,{\bm r})G(\tau,{\bm r};\tau',{\bm r}')+\frac{1}{2}\Pi(\tau',{\bm r}';\tau,{\bm r})D(\tau,{\bm r};\tau',{\bm r}')\right. \\
       & + \frac{g^2}{2}G(\tau,{\bm r};\tau',{\bm r}')G(\tau',{\bm r}';\tau,{\bm r})D(\tau,{\bm r};\tau',{\bm r}')  +  \frac{v^2}{2}G(\tau,{\bm r};\tau',{\bm r}')G(\tau',{\bm r}';\tau,{\bm r})\delta({\bm r}-{\bm r}') \\
        &  +  \left. \frac{g'^2}{2}G(\tau,{\bm r};\tau',{\bm r}')G(\tau',{\bm r}';\tau,{\bm r})D(\tau,{\bm r};\tau',{\bm r}')\delta({\bm r}-{\bm r}') \right]\,.
  \end{split} 
\end{equation}
Note that the spatially random couplings lead to an additional $\delta({\bm r} - {\bm r}')$ in their contributions arising from the disorder averages in (\ref{f5}) and (\ref{f7}). The saddle point of (\ref{Sall2}) leads to equations for the Green's functions which can also be derived from Fig.~\ref{fig:yukawa}:
\begin{align}
&\Sigma(\tau,{\bm r}) = g^2 D(\tau,{\bm r})G(\tau,{\bm r}) +  v^2 G(\tau,{\bm r}) \delta^2({\bm r}) +  {g'}^2G(\tau,{\bm r})D(\tau,{\bm r})\delta^2({\bm r}), \nonumber \\
& \Pi(\tau,{\bm r}) = -g^2 G(-\tau,-{\bm r})G(\tau,{\bm r}) -  {g'}^2G(-\tau,{\bm r})G(\tau,{\bm r})\delta^2({\bm r}), \nonumber \\
&G(i\omega,{\bm k}) = \frac{1}{i\omega-\varepsilon({\bm k})+\mu-\Sigma(i\omega,{\bm k})}, \nonumber \\
&D(i\Omega,{\bm q}) = \frac{1}{\Omega^2+{\bm q}^2+m_b^2-\Pi(i\Omega,{\bm q})}. 
\label{eq:saddle_pt_eqs}
\end{align}
Before discussing the solution of (\ref{eq:saddle_pt_eqs}), the computation of response functions of fermion bilinears, such as the conductivity, is described. These can be obtained by inserting external sources into (\ref{Sall2}) and then taking the variational derivatives with respect to them. This leads to the graphs shown in Fig.~\ref{fig:ladders}, which have to evaluated with fully renormalized Green's functions.

The following subsections discuss the solutions of the equations in (\ref{eq:saddle_pt_eqs}) and Fig.~\ref{fig:ladders} for the cases without and with spatial randomness.

\subsection{No spatial randomness}

The solution of (\ref{eq:saddle_pt_eqs}) with $g \neq 0$, but $g'=0$ and $v=0$, is considered. This corresponds to the quantum phase transition without disorder, and has been much studied in the literature.
At the quantum critical point, (\ref{eq:saddle_pt_eqs}) yields a non-Fermi liquid form for the fermion Green's function, and a Landau-damped form for the boson Green's function \cite{PALee89,Patel1}
\begin{align}
\Sigma (i \omega, {\bm k}) \sim -i \mbox {sgn} (\omega) |\omega|^{2/3} \,, & \quad G( i\omega,{\bm k}) = \frac{1}{i \omega - \varepsilon({\bm k}) - \Sigma (i \omega, {\bm k})} \nonumber \\ \quad D(i \Omega,{\bm q} ) & = \frac{1}{\Omega^2 + {\bm q}^2 + \gamma |\Omega|/q}\,, \label{g1}
\end{align}
The fermion Green's function has a sharp Fermi surface in momentum space, and ${\bm k}$ in (\ref{g1}) is assumed to be close to the Fermi surface. But $G$ is diffusive in frequency space, indicating the absence of well-defined fermionic quasiparticles.

However, an important point is that essentially none of this non-Fermi liquid structure feeds into the conductivity, which remains very similar to that of a Fermi liquid \cite{Maslov12,Maslov17a,Maslov17b,Guo2022,SenthilShi22,GuoIII} with the form:
\begin{align}
\sigma (\omega) \sim \frac{1}{-i \omega} + |\omega|^0 + \cdots \quad \mbox{($\omega^{-2/3}$ term has vanishing co-efficient)} \label{g2}
\end{align}
There has been a claim \cite{YBK94} of a $\omega^{-2/3}$ contribution to the conductivity, but its co-efficient vanishes after evaluation of all the graphs in Fig.~\ref{fig:ladders} \cite{SenthilShi22,Guo2022}. This cancellation can be understood as a consequence of Kohn's theorem \cite{Kohn61}, which states that in a Galilean-invariant system only the first term of the right-hand-side of (\ref{g2}) is non-zero.
A Galilean-invariant system is not considered here, but all contributions to the possible $\omega^{-2/3}$ term arise from long-wavelength processes in the vicinity of patches of the Fermi surface, and these patches can be embedded in a system which is Galilean-invariant also at higher energies.

\subsection{With spatial randomness}

The absence of the $\omega^{-2/3}$ term in (\ref{g2}) is a strong indication that the resolution of the strange metal problem cannot come from a clean system. There can be umklapp processes which dissipate momentum, but these require special features of the Fermi surface to survive at low momentum. A theory with only potential scattering disorder as in (\ref{f4}) and (\ref{f5}), {\it i.e.} $g \neq 0$, $v \neq 0$, but $g'=0$, is also not sufficient \cite{Guo2022,Foster2022}: it leads to marginal Fermi-liquid behavior in the electron self energy, but no strange metal behavior in transport.
So for a generic and universal theory of strange metals, the influence of disorder with $g$, $g'$, and $v$ all non-zero should be considered.

The solution of (\ref{eq:saddle_pt_eqs}) yields a boson Green's function which has a diffusive form at the critical point \cite{HLR} 
\begin{align}
D(q, i \Omega) \sim \frac{1}{q^2 + \gamma |\Omega|} \label{g3}
\end{align}
Inserting this into the fermion Green's function gives a marginal Fermi liquid form \cite{HLR, Patel2}
\begin{align}
G (\omega) & \sim  \frac{1}{ \displaystyle 
\omega \, \frac{m^\ast (\omega)}{m} - \varepsilon ({\bm k}) + 
i \left(\frac{1}{\tau_e} + \frac{1}{\tau_{\rm in} (\omega)} \right) \mbox{sgn} (\omega) } \nonumber \\
 \frac{1}{\tau_{e}} \sim v^2 \quad ; \quad & \frac{1}{\tau_{\rm in} (\omega)} \sim \left( \frac{g^2}{v^2} + g^{\prime 2} \right) |\omega | \quad ; \quad  
\frac{m^\ast (\omega)}{m} \sim \frac{2}{\pi} \left( \frac{g^2}{v^2} + g^{\prime 2} \right) \ln (\Lambda/\omega)  \label{g4}
\end{align}
The expressions in the second line are schematic, and show only the dependence upon $g$, $g'$ and $v$ without numerical constants.
This result matches the photoemission observations in (\ref{p3}) for $\alpha = 1/2$. Note that there are two distinct contributions to the singular $|\omega|$ electron inelastic scattering rate $1/\tau_{\rm in}$: one from the combination of impurity scattering $v$ with the spatially uniform interaction $g$ \cite{HLR}, and the other from the spatially random interaction $g'$ \cite{Altman1,Patel2}.

Inserting these solutions for the Green's functions into the action in (\ref{Sall2}), gives a $T \ln (1/T)$ specific heat \cite{Patel1}.

Turning to the evaluation of the conductivity graphs in Fig.~\ref{fig:ladders}, the key property of the strange metal, the conductivity, is given by \cite{Altman1,Patel2}
\begin{align}
\sigma (\omega) & \sim \frac{1}{ \displaystyle \frac{1}{\tau_{\rm trans} (\omega)} - i \omega \, \frac{m_{\rm trans}^\ast (\omega)}{m}} \\
 \frac{1}{\tau_{\rm trans} (\omega)} \sim v^2 & + g^{\prime 2} |\omega|  \quad; \quad 
\frac{m_{\rm trans}^\ast (\omega)}{m} \sim \frac{2g^{\prime 2}}{\pi} \ln (\Lambda/\omega) \label{g5}
\end{align}
This expression shows that the residual resistivity $\rho_0$ at $T=0$ is determined by the elastic scattering rate $1/\tau_e \sim v^2$,  as in a disordered Fermi liquid. The inelastic processes lead to a frequency and temperature dependence which matches precisely with the observational form in (\ref{p2}). An important feature is that of the two processes contributing to the electron inelastic scattering rate $1/\tau_{\rm in}$ in (\ref{g4}), only one contributes to the inelastic transport rate $1/\tau_{\rm trans}$. The processes involving the spatially uniform interaction $g$ and the impurity potential $v$ {\it cancel out\/} in the computation of the conductivity from Fig.~\ref{fig:ladders}, and {\it only\/} those involving the spatially random interaction $g'$ survive \cite{Patel2}. A consequence of this cancellation is that the constant $\alpha$ in (\ref{Planckian}) approaches $\alpha = \pi/2$ for the quasiparticle $m^\ast$ in the limit $g' \gg g$ \cite{Patel1}, and decreases from this value as $g$ is increased \cite{Patel2}.

To summarize, the conductivity of the theory $\mathcal{L}_c + \mathcal{L}_v + \mathcal{L}_\phi$ yields the strange metal conductivity in (\ref{p1}), 
with $\rho_0 \sim v^2$ and $A \sim g^{\prime 2}$. Note that the value of $g$ does not make a direct difference to the value of the linear-$T$ resistivity, although it does affect the marginal Fermi liquid behavior of the electron self energy, as noted in (\ref{g4}). It is also notable that the residual resistivity and linear-$T$ resistivity slope are determined by different sources of disorder: those in (\ref{f5}) and (\ref{f7}) respectively. This distinction should be important in understanding trends in observations \cite{Ong91,IISC94}.

The key role of spatial randomness in the Yukawa coupling in this theory implies a prediction: correlated electron systems will not exhibit low $T$ strange metal behavior in sufficiently clean samples. Evidence in support of this prediction has appeared in recent experiments on graphene: while twisted bilayer graphene has a strange metal phase \cite{CaoStrange}, the much cleaner system of rhombohedral trilayer graphene does not \cite{YoungNotStrange}.

Finally, note that a recent computation \cite{PatelNoise} of shot noise in the $g'$-$v$ model yields results in agreement with observations \cite{NatelsonNoise}.

\subsection{Fermi volume changing transitions}
\label{sec:kondo}

So far the theory in Section~\ref{sec:univ} in which the critical boson $\phi$ represented the order parameter for the Ising-nematic transition in Fig.~\ref{fig:nematic} has been considered. It is evident \cite{Metlitski1} that the same theory applies to other order parameters at zero momentum {\it e.g.\/} ferromagnetism. A somewhat less trivial statement is that not much changes for order parameters at non-zero momentum {\it e.g.\/} antiferromagnetism or spin-density-wave order. The important coupling $g'$ is spatially random, and so momentum is not conserved at the Yukawa interaction vertex: consequently the singular structure remains unchanged for order parameters at non-zero momentum. Furthermore, the theory applies even when the $c$ fermions have a three-dimensional dispersion, as long as the spin fluctuations remain quasi-two-dimensional, as is the case in YbRh$_2$Si$_2$ \cite{Steglich00}.

\begin{figure}
\begin{center}
\includegraphics[width=5.2in]{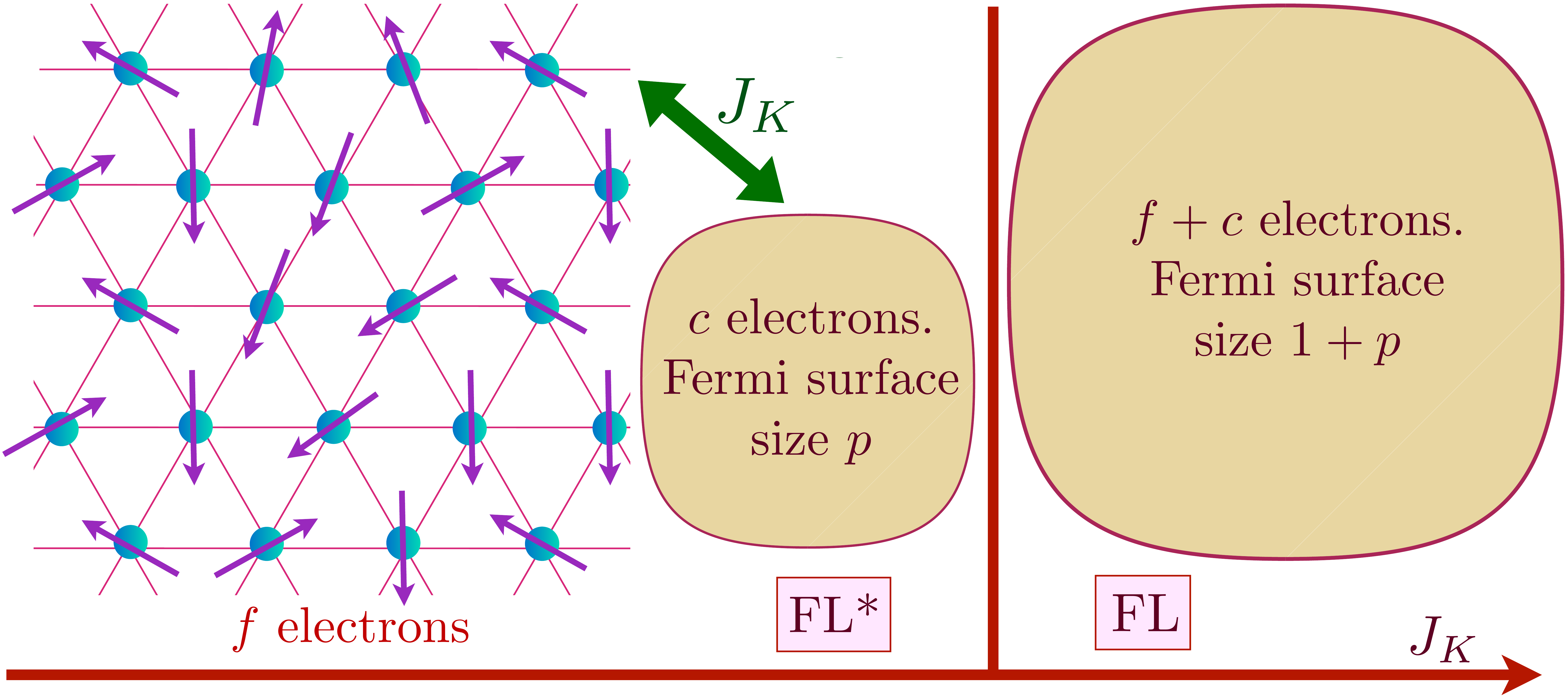}
\end{center}
\caption{Fermi volume changing transition in a Kondo lattice of spins of $f$ electrons and a conduction band of $c$ electrons. In the fractionalized Fermi liquid (FL*) phase the $f$ electrons form a spin liquid with fractionalized spinon excitations, while the $c$ electrons form a `small' Fermi surface, In the FL phase, the $f$ and $c$ electrons hybridize, and realize a `large' Fermi surface which has the Luttinger volume of free electrons.}
\label{fig:kondo}
\end{figure}

Indeed, the universal theory of Section~\ref{sec:univ} applies {\it also\/} to quantum transitions without a symmetry-breaking order parameter. The most experimentally relevant cases of such transitions are those that involve a change in the volume of the Fermi surface. 
For the Kondo lattice, there can be a transition \cite{SVS03,SVS04} from a fractionalized Fermi liquid (FL*) to a Fermi liquid (FL) as illustrated in Fig.~\ref{fig:kondo}. In this case, the theory of Section~\ref{sec:univ} applies \cite{Altman1} after $\phi$ is chosen as  the complex (`slave') boson $\phi \sim f_{\alpha}^\dagger c_\alpha$, where $\alpha$ is a spin index. The FL phase is obtained when $\phi$ is condensed, and the FL* phase appears when there is no $\phi$ condensate. A recent example of a FL*-FL transition in a Kondo lattice system 
appears in the observations of Maksimovic {\it et al.\/} \cite{Analytis22} in CeCoIn$_5$.
\begin{figure}
\begin{center}
\includegraphics[width=6.1in]{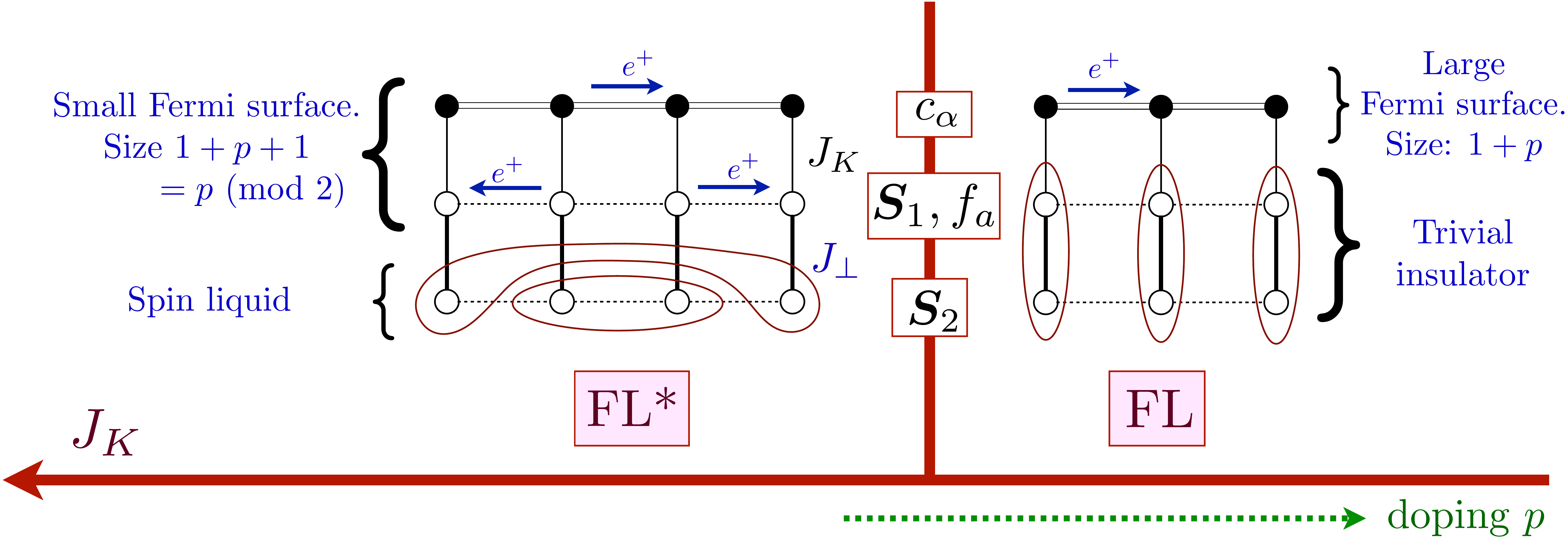}
\end{center}
\caption{Fermi volume changing transition in the ancilla theory of the Hubbard model. For $J_\perp \gg J_K$, the ancilla ${\bm S}_{1,2}$ spins form rung singlets, and a FL phase is present in the $c_\alpha$ layer. For $J_K \gg J_\perp$, we obtain the FL* phase in which the $c_\alpha$ electrons hybridize with the ${\bm S}_1$ spins to form a Fermi surface similar to the large Fermi surface of the Kondo lattice model in Fig.~\ref{fig:kondo}, while the ${\bm S}_2$ spins form a spin liquid with neutral spinon excitations.}
\label{fig:ancilla}
\end{figure}

For the hole-doped cuprates, a Fermi-volume changing transition has been proposed \cite{YaHui-ancilla1,YaHui-ancilla2} in which the FL* phase realizes hole pockets observed in the pseudogap metal \cite{Ramshaw22}. 
A theory for the FL*-FL transition in a single band Hubbard-like model was obtained by a canonical mapping of the Hubbard model to a 3-layer model in which free electrons ($c_\alpha$) are coupled via a Kondo coupling $J_K$ to 
a bilayer antiferromagnet of $S=1/2$ ancilla spins ${\bm S}_{1,2}$ with interlayer exchange $J_\perp$ (see Fig.~\ref{fig:ancilla}).
Eliminating the ancilla spins in a large $J_\perp$ expansion via a Schrieffer-Wolff transformation \cite{QPMbook} yields a Hubbard model for the $c_\alpha$ layer with $U \sim J_K^2/J_\perp$ \cite{Mascot22}. As shown in Fig.~\ref{fig:ancilla}, a FL*-FL transition is obtained in this model, with the FL* phase present at large $J_K$. Note the inversion from the Kondo lattice model in Fig.~\ref{fig:kondo}, where the FL* phase was obtained at small $J_K$. A theory for the FL*-FL transition now requires a multicomponent complex boson $\Phi_{\alpha a} \sim f_{a}^\dagger c_\alpha$, where the $f_a$ are `ghost' fermions representing the ${\bm S}_1$ spins, and $a$ is a SU(2) gauge index \cite{YaHui-ancilla1,YaHui-ancilla2}. Continuing the inversion with 
the Kondo lattice model in Fig.~\ref{fig:kondo}, the FL* phase is now obtained when $\Phi_{\alpha a}$ is condensed, while the FL phase is obtained otherwise. But the critical theory is very similar to that of the Kondo lattice model, and a strange metal phase is obtained, as in Ref.~\cite{Altman1}, when $\Phi_{\alpha a}$ has a spatially random coupling to the fermions.
This model can also explain the observed peak in the specific heat \cite{Michon19} in the critical region by the specific heat of the $f_a$ ghost fermions \cite{YaHui-ancilla1}.

In the application for the cuprates, it is believed that the FL* phase is ultimately not stable down 
to $T=0$ \cite{YaHui-ancilla1,Christos:2023oru,CS23}, 
and so there is no actual FL*-FL quantum critical point. Nevertheless, the theory of such a quantum critical point can still be relevant for the strange-metal behavior in the quantum-critical region at non-zero temperatures, as noted in Fig.~\ref{fig:nematic}.

An alternative approach to the pseudogap metal of the cuprates is to view it as a metal of holons {\it i.e.\/} charge $+e$ but spinless fermions (in contrast to the spin-1/2 fermions in the FL* phase). There are reasonable comparisons of the holon metal theory to experimental data \cite{HeScheurer19} and numerical studies \cite{WuScheurer1,Scheurer:2017jcp,WuScheurer2}.
The above considerations on Fermi volume changing transitions apply also to theories \cite{sdw09,DCSS15b,DCSS15,WuScheurer1,Scheurer:2017jcp,Sachdev:2018ddg,WuScheurer2,Bonetti22} of the transition from the holon metal to the Fermi liquid.

\section{Quantum black holes and holography}
\label{sec:qbh}

Connections between the SYK model and black holes are reviewed more completely in Ref.~\cite{Sachdev23}; a brief discussion is presented here.

Formally, the quantum theory of gravity and electromagnetism can be written as a Feynman path integral over the spacetime metric $g_{\mu\nu}$, and the electromagnetic gauge field $a_\mu$: this involves computing the partition function
\begin{equation}
\mathcal{Z} = \int \mathcal{D} g_{\mu \nu} \mathcal{D} a_\mu \exp \left( - \frac{1}{\hbar} \int d^d x \int_0^{\hbar/(k_B T)} \!\!\!\!\!\! d \tau 
\sqrt{g} \, \mathcal{L}_d [ g_{\mu\nu}, a_\mu ] \right) \label{feyn}
\end{equation}
over fields in $(d+1)$-dimensional spacetime, with $\mathcal{L}_d$ the Lagrangian of classical Einstein-Maxwell theory in $d+1$ spacetime dimensions, and $g$ the determinant of the metric. Here $\tau$ is time analytically continued to the imaginary axis, which is taken to lie on a circle of circumference $\hbar/(k_B T)$, where $\hbar$ is Planck's constant. 
This constraint on imaginary time follows from the correspondence between the evolution operator ${U}(t)$ for real time $t$ in quantum mechanics, and the Boltzmann-Gibbs partition function $\mathcal{Z}$ for a quantum system with Hamiltonian $\mathcal{H}$:
\begin{equation}
{U} (t)  = \exp \left( - i \mathcal{H} t/ \hbar \right) \quad \Leftrightarrow \quad \mathcal{Z} = \mbox{Tr} \exp \left( - \mathcal{H}/(k_B T) \right)\,.
\end{equation}

\begin{figure}
\begin{center}
\includegraphics[width=3.2in]{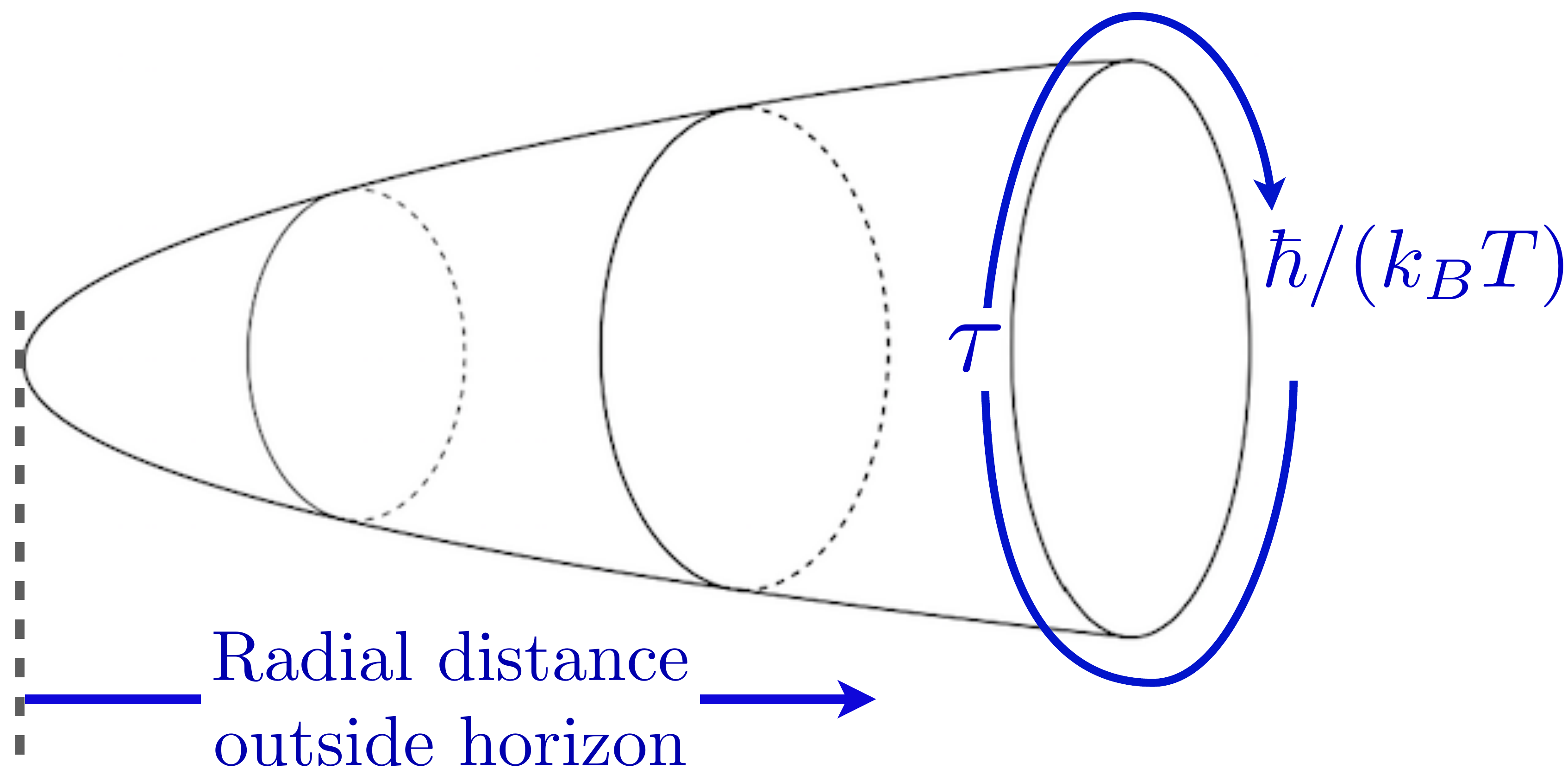}
\end{center}
\caption{Spacetime geometry outside a black hole. Only the radial direction and the imaginary time direction $\tau$ are shown, and the two angular directions are not shown.}
\label{fig2}
\end{figure}

The fate of quantum gravity near black-hole saddle-points is of interest here. The formation of a black hole requires dense matter, but it is assumed that this matter has collapsed to the center of the black hole, so that only the spacetime away from this singularity is of interest. Then the matter-free theory in (\ref{feyn}) can be assumed, and the mass of the black hole only appears as a boundary condition on the gravitational field. 
Similar comments apply to the charge of the black hole, which leads to a familiar Gauss-law condition on the electromagnetic field.
The electrically neutral (Schwarzschild) and electrically charged (Reissner-N\"ordstrom) solutions of the Einstein-Maxwell equations are obtained as saddle-points of (\ref{feyn}), and analytically continued to imaginary time. Then the
spacetime geometry outside a black hole turns out to be that of a `cigar' as shown in Fig.~\ref{fig2}.
Note that all dependence of (\ref{feyn}) on $\hbar$ and $T$ is explicit, and there is no $\hbar$ or $T$ in the Lagrangian $\mathcal{L}_d$.

The path integral in (\ref{feyn}) is pathological because it includes infinities that cannot be controlled by the usual renormalization tricks of quantum field theory. Nevertheless, Gibbons and Hawking \cite{Gibbons_Hawking} boldly decided to evaluate it in the semi-classical limit, where they only included the contribution of the cigar saddle point outside the horizon in Fig.~\ref{fig2}. For a black hole, they also imposed the requirement that spacetime was smooth at the horizon in imaginary time. From this relatively simple computation, they were able to obtain the thermodynamic properties of a black hole, including its temperature and entropy. For a neutral black hole of mass $M$ in $d=3$ they found
\begin{equation}
\frac{S}{k_B} =  \frac{c^3 A}{4 \hbar G} \quad, \quad \frac{k_B T}{\hbar} = \frac{c^3}{8 \pi G M} \label{ST}
\end{equation}
where $c$ is the velocity of light, $G$ is Newton's gravitational constant, and $A = 4 \pi R^2$ is the {\it area} of the black hole horizon with $R = 2 GM/c^2$ the horizon radius. Hawking also discussed the eventual evaporation of such black holes, but the discussion here is limited to times shorter than the inverse evaporation rate. 

The revolutionary results in (\ref{ST}) raised many more questions than they answered. Is this semi-classical computation of thermodynamics compatible with Boltzmann's fundamental statistical interpretation of entropy in (\ref{b2})? How does a computation in imaginary time outside a black hole know about the entropy of quantum degrees of freedom inside a black hole? Can the energy eigenvalues of a quantum Hamiltonian describing the inside of a black hole whose density of states $D(E)$ yields a $S(E)$ that is consistent with (\ref{ST}), and the partition function $\mathcal{Z}$ in (\ref{feyn}), be computed? With the energy $E$ shifted so that $E=0$ for the ground state, $\mathcal{Z}$ is related to $D(E)$ by
\begin{equation}
\mathcal{Z} = \int_{0^-}^\infty dE D(E) \exp\left( - \frac{E}{k_B T} \right) \, . \label{ZD}
\end{equation}
Many other questions are raised in considering the fate of the black hole as it evaporates while emitting blackbody radiation at the temperature in (\ref{ST}), and computes the entanglement entropy of the Hawking radiation.

\begin{figure}
\begin{center}
\includegraphics[width=3.2in]{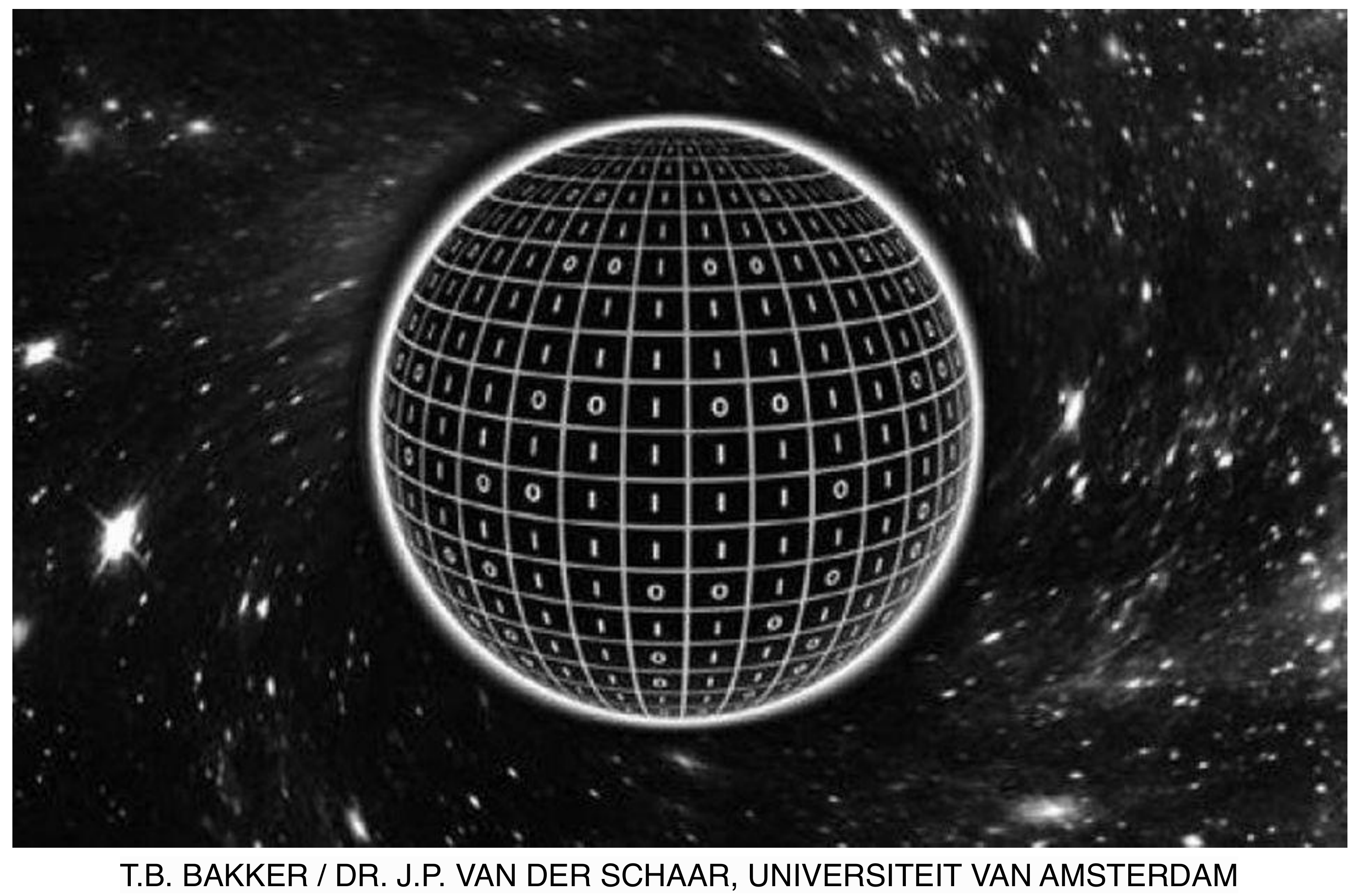}
\end{center}
\caption{Holography: the number of qubits required for the quantum simulation of a black hole is proportional to its surface area.}
\label{fig3}
\end{figure}

A remarkable feature of the entropy in (\ref{ST}) is that it is proportional to the surface {\it area\/} of the black hole. This contrasts with extensive volume proportionality of the entropy, mentioned below (\ref{b1}), obeyed by all other quantum systems. Attempts to understand this feature led to the idea of holography \cite{tHooft,Susskind,Maldacena} illustrated in Fig.~\ref{fig3}. 
It is proposed that a quantum simulation of a black hole can be built by simple two-level systems {\it i.e.\/} qubits. How many qubits are needed? As $N$ qubits can describe $2^N$ linearly-independent quantum states, 
the formula in (\ref{b2}) implies that the number of qubits is proportional to the entropy, and hence the area of the black hole. So the qubits realize a many-body quantum system which can be thought of as residing on its surface {\it i.e.\/} the qubits are a faithful $(d-1)$-dimensional hologram of the complete quantum gravitational theory of the black hole in $d$ spatial dimensions.  

A specific realization of a quantum simulation was found in string theory for `extremal' black holes (defined below) with low-energy supersymmetry \cite{David02}. This realization has a ground state with an exponentially large degeneracy, and special features of the supersymmetry were employed to compute this degeneracy, yielding
\begin{equation}
D(E) = \left[\exp (S(E)/k_B)\right] \delta(E) + \ldots\,. \label{Ddelta}
\end{equation}
where $\ldots$ refers to a continuum above an energy gap.
The value of $S(0)$ was found to be precisely that in the Hawking formula in (\ref{ST}) \cite{Strominger96}. However, the zero energy delta-function in (\ref{Ddelta}) is known \cite{matt22,luca22b} to be a special feature of theories with low energy supersymmetry, and is not a property of the generic semi-classical path integral over Einstein gravity in (\ref{feyn}), as 
discussed below (see Fig.~\ref{fig5}).

To move beyond theories with low energy supersymmetry, any general constraints that must be obeyed by the many-body system realized by the interactions between the qubits must be considered. An important constraint comes from an earlier result by Vishveshwara \cite{cvc}. He computed the relaxation time, $t_r$ of quasi-normal modes of black holes in Einstein's classical theory; this is the time in which a black hole relaxes exponentially back to a spherical shape after it has been perturbed by another body:  
\begin{equation}
t_r = \alpha' \frac{8 \pi G M }{c^3} \,,\label{tr}
\end{equation}
where $\alpha'$ is a numerical constant of order unity dependent upon the precise quasi-normal mode.
Comparing Vishveshwara's result in (\ref{tr}) with Hawking's result in  (\ref{ST}), implies
\begin{equation}
t_r = \alpha' \frac{\hbar}{k_B T} \label{tra}
\end{equation}
which is exactly of the form in (\ref{taueq}) for many-body quantum systems without quasiparticles! This is a key hint that the holographic quantum simulation of a black hole must involve a quantum system without quasiparticle excitations, if it is to reproduce basic known features of black-hole dynamics. At this point, it is interesting to note that measurements of $t_r$ in binary black-hole mergers by LIGO-Virgo \cite{BHbound} do indeed fall around the value of $\hbar/(k_B T)$.

Can insights from the path integral over time reparameterizations of the SYK model in (\ref{feynsyk}) be used to evaluate the path integral over spacetime metrics of black holes in (\ref{feyn})? Remarkably, for a black hole with a non-zero fixed total charge ${Q}$, the answer is yes, as is discussed in more detail in the companion article \cite{Sachdev23}.
The saddle-point solution of the Einstein-Maxwell action for a charged black hole has a spacetime which is 3+1 dimensional flat Minkowski far from the black hole, but it factorizes into a 1+1 dimensional spacetime involving the radial direction, and a 2-dimensional space of non-zero angular-momentum modes around the spherical black hole. 
As the black-hole temperature $T \rightarrow 0$ (also known as the extremal limit), the non-zero angular-momentum modes become unimportant, and the partition function of the charged black hole can be written purely as a theory of quantum gravity in 1+1 spacetime dimensions, which is an extension of a theory known as Jackiw-Teitelboim (JT) gravity \cite{JTrev}; then (\ref{feyn}) reduces to
\begin{equation}
\mathcal{Z}_{JT} = e^{A_0 c^3/(4 \hbar G)} \int \mathcal{D} g_{\mu \nu} \mathcal{D} a_\mu \mathcal{D} \Phi \exp \left( - \frac{1}{\hbar} \int d \zeta \int_0^{\hbar/(k_B T)} \!\!\!\!\!\! d \tau 
\sqrt{g} \, \mathcal{L}_{JT} [ g_{\mu\nu}, a_\mu, \Phi ] \right)\,, \label{feynQ}
\end{equation}
where all fields are in 1+1 spacetime dimensions, the Lagrangian $\mathcal{L}_{JT}$ additionally involves a scalar field $\Phi$ representing quantum fluctuations of the radius of the black-hole, and $A_0 = 2 G Q^2/c^4$ is the area of the black-hole horizon at $T=0$. The $(1+1)$-dimensional spacetime saddle point of $\mathcal{Z}_{JT}$ has a uniform negative curvature: it is the anti-de Sitter space AdS$_2$. Quantum gravity in 1+1 dimensions is especially simple because there is  no graviton, and it is possible to make an explicit holographic mapping to a quantum system in 0+1 dimensions. It turns out that the holographic quantum realization of the 1+1 dimensional theory $\mathcal{Z}_{JT}$ in (\ref{feynQ}) is exactly the $0+1$ dimensional SYK model partition function in $\mathcal{Z}_{SYK-TR}$ in (\ref{feynsyk}). 
The fluctuations of the metric in the boundary region between the 1+1 dimensional and 3+1 dimensional spacetimes 
are described by the time reparameterization $f(\tau)$, and the boundary value of $a_\mu$ becomes the phase field $\phi (\tau)$. 

This mapping from the original quantum-gravity path integral in (\ref{feyn}) for charged black holes to the SYK path integral in (\ref{feynsyk}) is very powerful.
It enables progress beyond the semi-classical results of Hawking in (\ref{ST}) for black holes with non-zero charge ${Q}$. Applying this mapping from $\mathcal{Z}_{JT}$ to $\mathcal{Z}_{SYK-TR}$, a density of states $D(E)$ with precisely the $E$ dependence in (\ref{de}) is obtained, which also corresponds to that shown for the SYK model in Fig.~\ref{fig5}a \cite{luca22a}.
\begin{equation}
D(E) \sim \left( \frac{A_0 c^3}{\hbar G} \right)^{-347/90} \exp\left( \frac{A_0 c^3}{4 \hbar G} \right) \sinh \left( \left[\sqrt{\pi} A_0^{3/2} \frac{c^3}{\hbar G} \frac{E}{\hbar c} \right]^{1/2} \right) \,. \label{DEF}
\end{equation}
In contrast to the supersymmetric result in (\ref{Ddelta}) and Fig.~\ref{fig5}b \cite{luca20,matt22}, the non-supersymmetric result in (\ref{DEF}) has no delta function at zero energy, and so no exponentially large ground-state degeneracy. 
The exponential factor in (\ref{DEF}) is related to the Bekenstein-Hawking black-hole entropy in (\ref{ST}), while the exponential factor in (\ref{de}) is the SYK zero-temperature entropy \cite{GPS2}. The $\sinh$ factors are identical between (\ref{de}) and (\ref{DEF}), and the co-efficient of $E$ is determined from the linear-in-$T$ dependence of the entropy in (\ref{SSYK}) for the SYK model, and from 
the linear-in-$T$ dependence also found in the Bekenstein-Hawking black-hole entropy. The prefactors of a power of $A_0$ in (\ref{DEF}) and of $1/N$ in (\ref{de}) are not universal, and require knowledge of the underlying matter content \cite{GKST,luca22a}.

\section{Conclusions}

The SYK model originally appeared as a solvable model of many interacting quantum particles which does not have any particle-like excitations. The efforts of theorists across the condensed matter and quantum gravity communities over several decades have led to a fairly complete and deep understanding of its properties, as was reviewed in Section~\ref{sec:SYK}. The SYK model relaxes rapidly to thermal equilibrium in the Planckian time $\hbar / (k_B T)$. It has an exponentially large number of very low energy many-body eigenstates, with wavefunctions which change chaotically from one state to the next (and are not related by the addition or removal of a small number of quasiparticles).

But with infinite-range interactions, and no spatial structure, the SYK model initially appears rather artificial, and with no direct connection to any realistic physical system. 

Section~\ref{sec:strange} described extensions of the SYK model, by adding both fermionic and bosonic degrees of freedom, a spatial structure, and a Fermi surface for the fermions. These extensions are directly connected to microscopic models of experimentally relevant quantum phase transitions, both with and without symmetry-breaking order parameters. They yield a universal theory of the strange-metal state ubiquitous in quantum materials with strong electron-electron interactions. A key feature of this universal theory is that the Yukawa amplitude for a fermion to emit or absorb a boson varies randomly from point to point in space (but is time-independent). Highlights of experimental observations on strange metals were presented in the beginning of Section~\ref{sec:strange}, and the physical properties of the universal theory are consistent with all of them. 

More surprisingly, the SYK model of Section~\ref{sec:SYK}, without any extensions, provides a faithful quantum simulation of the low temperature properties of charged black holes in 3 space dimensions. Charged black hole solutions of  Einstein's theory of gravity and Maxwell's theory of electromagnetism have the striking property that the spacetime metric factorizes near the horizon of the black-hole. One factor represents the two angular spatial directions at a fixed radius outside the black-hole: this factor can be safely ignored in the low energy theory. The other factor is a two-dimensional spacetime, representing the single radial direction, along with time. As summarized in Section~\ref{sec:qbh} and Ref.~\cite{Sachdev23}, remarkably, the theory of quantum gravity in this two-dimensional spacetime co-incides with the low energy theory of the SYK model of Section~\ref{sec:SYK}. This connection has allowed computations of many properties of quantum charged black holes, beyond those that were accessible in Hawking's semi-classical theory. In particular, it is now understood that the low energy density of quantum states of non-supersymmetric black holes resembles that of the SYK model shown in Fig.~\ref{fig5}a. Only supersymmetric black holes, and supersymmetric SYK models, have an exponentially large exact degeneracy in their ground states, as sketched in Fig.~\ref{fig5}b.

\subsection*{Acknowledgements} I thank Aavishkar Patel, Haoyu Guo and Ilya Esterlis for collaboration on the work described in Section~\ref{sec:strange}. This research has been supported by the U.S. National Science Foundation, most recently under grant No. DMR-2245246.

\bibliography{refsore}

\end{document}